\documentclass[aps,prd,preprint,superscriptaddress,showpacs,showkeys]{revtex4-1}
\pdfoutput=1
\usepackage[intlimits]{amsmath}
\usepackage{amssymb}
\usepackage[config=altsf]{subfig}
\usepackage{graphicx}% Include figure files
\usepackage{dcolumn}% Align table columns on decimal point
\usepackage{bm}% bold math
\usepackage{feynmf}
\begin{document}

\title{Leptophilic Dark Matter\\ and the Anomalous Magnetic Moment of the Muon}

\author{Prateek Agrawal}
\affiliation{Fermilab, P.O. Box 500, Batavia, IL 60510, USA}
\author{Zackaria Chacko}
\affiliation{Maryland Center for Fundamental Physics, Department of Physics, University of Maryland, College Park, MD 20742-4111 USA}
\author{Christopher B. Verhaaren}
\email[]{cver@umd.edu}
\affiliation{Maryland Center for Fundamental Physics, Department of Physics, University of Maryland, College Park, MD 20742-4111 USA}
\date{\today}

\begin{abstract}

We consider renormalizable theories such that the scattering of dark
matter off leptons arises at tree level, but scattering off nuclei
only arises at loop. In this framework, the various dark matter
candidates can be classified by their spins and by the forms of their
interactions with leptons.  We determine the corrections to the
anomalous magnetic moment of the muon that arise from these
interactions. We then consider the implications of these results for a
set of simplified models of leptophilic dark matter. When a dark
matter candidate reduces the existing tension between the standard
model prediction of the anomalous magnetic moment and the experimental
measurement, the region of parameter space favored to completely
remove the discrepancy is highlighted. Conversely, when agreement is
worsened, we place limits on the parameters of the corresponding
simplified model.  These bounds and favored regions are compared
against the experimental constraints on the simplified model from
direct detection and from collider searches.  Although these
constraints are severe, we find there do exist limited regions of
parameter space in these simple theories that can explain the observed
anomaly in the muon magnetic moment while remaining consistent with
all experimental bounds.

\end{abstract}


%muons, properties of/Dark Matter/radiative corrections EM/rad. corr. Electroweak

\maketitle

\section{Introduction\label{sec:intro}}

The standard model (SM) of particle physics is in very good agreement 
with experiment~\cite{PDG2012}. The recent discovery of a SM-like Higgs 
particle at the Large Hadron Collider (LHC)~\cite{CMS2012,Atlas2012} provides further 
confirmation that the SM is an excellent description of the 
interactions of the elementary particles up to the weak scale. 
Nevertheless, there is now compelling experimental evidence that
points to  physics beyond the SM. Among the phenomena that cannot 
be accounted for within the SM are the non-zero values of the neutrino 
masses, and the existence of a baryon asymmetry.

As cosmological data has become more precise, it has become clear that 
more than 20\% of the energy density in the universe is associated with 
some form of dark matter (DM)~\cite{Planckxvi2013}. While very little is 
known about the precise nature of DM, a large class of well motivated 
theories involve particles that have interactions of weak scale strength 
with visible matter. These include theories where DM is composed of 
WIMPs, particles of weak scale mass that survive as thermal relics
\cite{Lee:1977ua,Vysotsky:1977pe}, and 
some models of Asymmetric Dark Matter~\cite{Nussinov:1985xr,Gelmini:1986zz,Spergel:1984re,Barr:1990ca,Barr:1991qn,Kaplan:1991ah,Gudnason:2006ug,Gudnason:2006yj,Kitano:2004sv,Kitano:2005ge,Kaplan:2009ag} (see \cite{Zurek:2013wia} for a
recent review and references).

Several types of experiments are involved in the search for DM. These 
include direct detection experiments searching for DM scattering
events,
%~\cite{
%Benetti:2007cd,
%Ahmed:2009zw,
%Bernabei:2010mq,
%Akimov:2011tj,
%Angloher:2011uu,
%Armengaud:2012pfa,
%Aalseth:2012if,
%Kim:2012rza,
%Behnke:2012ys,
%Baudis:2012zs,
%Agnese:2013rvf,
%Akerib:2013tjd}, 
indirect detection experiments searching 
for the products of DM annihilation such as photons and neutrinos,
%~\cite{
%Barwick:1997ig,
%Desai:2004pq,
%Aguilar:2007yf,
%Chang:2008aa,
%FermiLAT:2011ab,
%Aartsen:2012kia,
%fortheFermiLAT:2013naa,
%Adriani:2013uda,
%Fermi-LAT:2013uma,
%Ackermann:2013yva}, 
and colliders
%~\cite{
%Abbiendi:2000hh,
%Heister:2002ut,
%Abazov:2003gp,
%Abdallah:2003np,
%Achard:2003tx,
%Abdallah:2008aa,
%Aad:2011xw,
%Aaltonen:2012ek,
%Aaltonen:2012jb,
%ATLAS:2012ky,
%Chatrchyan:2012me,
%Chatrchyan:2012tea,
%Aad:2013oja} 
such as the LHC that seek to produce DM. 
Over the years these experiments have become increasingly sensitive, and 
their limits now exclude a significant part of the preferred parameter 
space for many WIMP DM candidates.

With the WIMP paradigm beginning to come under strain, several ideas 
have been put forward to explain the absence of a signal in these 
experiments. Perhaps the simplest possibility is that DM is leptophilic, 
coupling preferentially to leptons rather than to 
quarks~\cite{
Krauss:2002px,
Baltz:2002we,
Hambye:2006zn,
Chen:2008dh,
Fox:2008kb,
Cirelli:2008pk,
Ibarra:2009bm,
Davoudiasl:2009dg,
Kopp:2009et,
Cohen:2009fz,
Spolyar:2009kx,
Goh:2009wg,
Cao:2009yy,
Ko:2010at,
Chao:2010mp,
Schmidt:2012yg,
Das:2013jca}. 
This can help explain the null results of each of the different classes of 
experiments. Since the scattering of DM off nuclei now arises only at 
loop, the constraints from direct detection are weaker~\cite{Kopp:2009et}. 
Because the LHC is a hadron collider, the production of particles that do 
not have significant couplings to quarks or gluons is suppressed, and the 
resulting limits are again relatively weak. The limits on leptophilic DM 
from indirect detection also tend to be less severe. Since the average 
number of photons produced in DM annihilation to electrons and muons is 
much smaller than in the case of hadronic final states, the constraints 
from continuum photons are in general weaker, for example 
\cite{Papucci:2009gd,Abazajian:2010sq,
Hutsi:2010ai,
Abazajian:2010zb,
Hooper:2011ti,
Blanchet:2012vq,
Hooper:2012sr,
Tavakoli:2013zva,
Cholis:2013ena,
Ackermann:2013yva}.
Limits on DM from gamma ray lines 
tend to be significantly weaker than the limits from continuum photons, as 
may be seen from model independent 
analyses~\cite{Goodman:2010qn,Abazajian:2011tk} and are therefore easily 
satisfied. Although constraints from positron flux measurements are 
stronger for annihilations into electrons and muons, scenarios where 
leptophilic DM emerges as a thermal relic remain viable  
\cite{Kyae:2009jt, Bi:2009uj, Haba:2010ag, 
Carone:2011iw, Bergstrom:2013jra, Ibarra:2013zia}.

In general, DM candidates that couple directly to SM leptons with weak
scale strength are expected to have a significant impact on the
anomalous magnetic moment of the muon,
$a_{\mu}$~\cite{Fukushima:2013efa, Kopp:2014tsa}. It may therefore be
possible to obtain a tighter constraint on leptophilic DM from the
existing precision measurements of $a_{\mu}$ than from the experiments
considered in the previous paragraph. Furthermore, with the reduction
of experimental and theoretical uncertainties in the determination of
$a_{\mu}$, a 3$\sigma$ discrepancy between the two has remained. The
discrepancy is given by $\Delta a_{\mu}\equiv
a_{\mu}^{\text{Exp}}-a_{\mu}^{\text{SM}} =28.7(6.3)(4.9)\times
10^{-10}$~\cite{ Kinoshita:2005sm, Czarnecki:2002nt, Hagiwara:2011af,
Davier:2004gb, Passera:2004bj, Bennett:2006fi, PDG2012}, where the
errors shown represent the experimental and theoretical uncertainties,
respectively.
The theoretical error is dominated by hadronic contributions~\cite{
Davier:2003pw, Blokland:2001pb, Melnikov:2003xd, deTroconiz:2004tr,
Bijnens:2007pz, Davier:2010nc} (further references and reviews can be
found in \cite{Miller:2012opa, Jegerlehner:2009ry}).  The total
uncertainty including  both the experimental and theoretical
contributions is given by $\delta a_{\mu}\equiv 8.0\times 10^{-10}$.
If this observed anomaly is a consequence of new physics, a very
interesting possibility is that it is associated with the interactions
of leptons with DM. It is therefore important to identify theories of
DM that can give rise to such a signal.

In this paper we consider the contributions to the anomalous magnetic 
moment of the muon in theories of leptophilic DM. We focus on 
renormalizable theories where the scattering of DM off leptons arises at 
tree level, but DM scattering off nuclei only arises at loop. We also 
assume that the interactions of DM respect $CP$ so that there are no 
corrections to the electric dipole moments of the leptons. In such a 
framework, the various DM candidates can be classified by their spins
and by the forms of their interactions with leptons. For each case, we 
determine the corrections to $a_{\mu}$ that arise from these 
interactions. We then apply these results to several specific simplified 
models of DM. For concreteness, we focus on theories where the DM 
candidate couples universally with equal strength to all 3 flavors of SM 
leptons. 

When the DM candidate reduces the tension between the predicted value of 
$a_{\mu}$ in the SM and the experimental measurement, we highlight the 
region of parameter space favored to completely remove the discrepancy. 
Conversely, when agreement with experiment is worsened, we place limits 
on the parameters of the corresponding simplified model.  These bounds 
and favored regions are compared against the experimental constraints on 
the simplified model from direct detection and collider experiments. We 
find that although these constraints are severe, there do exist limited 
regions of parameter space in these simple theories that can explain the 
anomaly in $a_{\mu}$ while remaining consistent with current experimental 
bounds.

In section~\ref{sec:diagrams} we begin by
calculating the leading one loop contributions to $a_{\mu}$ that arise 
from renormalizable theories of leptophilic DM. In 
section~\ref{sec:compbnds} we determine the bounds on this scenario from 
direct detection and in section~\ref{sec:LEP} from LEP. These results
are applied to specific 
simplified models of DM in section~\ref{sec:models}.

\section{Contributions to $a_\mu$}
\label{sec:diagrams}
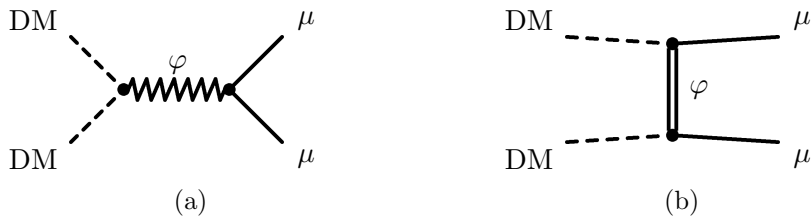
\begin{figure}[th]
\centering
\vspace{10mm}
\subfloat[ ]{
\centering
\begin{fmffile}{schan}
\begin{fmfgraph*}(100,40)\fmfkeep{schan}
\fmfpen{0.7thick}
\fmfleft{i1,i2}\fmfright{o1,o2}
\fmfv{l=DM}{i1}\fmfv{l=DM}{i2}\fmfv{l=$\mu$}{o1}\fmfv{l=$\mu$}{o2}
\fmf{dashes,tension=1}{i1,v1}\fmf{dashes,tension=1}{i2,v1}
\fmf{zigzag,label=$\varphi$,label.side=left,tension=1}{v1,v2}
\fmf{plain,tension=1}{v2,o1}\fmf{plain,tension=1}{v2,o2}
\fmfv{decor.shape=circle,decor.filled=full,decor.size=1.7thick}{v1}
\fmfv{decor.shape=circle,decor.filled=full,decor.size=1.7thick}{v2}
\end{fmfgraph*}
\end{fmffile}
\label{fig:schan}
}
\qquad \qquad \qquad
\subfloat[ ]{
\centering
\begin{fmffile}{tchan}
\begin{fmfgraph*}(100,40)\fmfkeep{tchan}
\fmfpen{0.7thick}
\fmfleft{i1,i2}\fmfright{o1,o2}
\fmfv{l=DM}{i1}\fmfv{l=DM}{i2}\fmfv{l=$\mu$}{o1}\fmfv{l=$\mu$}{o2}
\fmf{dashes,tension=1}{i1,v1}\fmf{dashes,tension=1}{i2,v2}
\fmf{double,label=$\varphi$,label.side=right,tension=0.15}{v1,v2}
\fmf{plain,tension=1}{v1,o1}\fmf{plain,tension=1}{v2,o2}
\fmfv{decor.shape=circle,decor.filled=full,decor.size=1.7thick}{v1}
\fmfv{decor.shape=circle,decor.filled=full,decor.size=1.7thick}{v2}
\end{fmfgraph*}
\end{fmffile}
\label{fig:tchan}
} \caption{Two possible channels of DM to muon scattering through a
mediator $\varphi$. In (a) $\varphi$ is electrically neutral and in
(b) it is charged.}\label{fig:chan}
\end{figure}

In this section we consider the contributions to $a_{\mu}$ from 
different theories of leptophilic DM. Our focus is on renormalizable 
theories where the scattering of DM off leptons arises at tree level. 
These theories can be separated into two broad categories depending on 
whether the particle that mediates the scattering of DM off leptons is 
neutral or electrically charged. These two classes of theories are shown 
in Fig.~\ref{fig:chan}.

These interactions can also affect the anomalous magnetic moment of the 
electron $a_{e}$. However, if the new physics couples universally 
to all the different flavors of leptons, its
contributions (henceforth denoted as $\hat{a}$)
to $a_e$ are much smaller than to $a_{\mu}$ and 
are therefore expected to easily satisfy the experimental bounds. The
one important exception to this statement is the case of very light
neutral mediators. Indeed, in analogy with QED, the one-loop
contribution to the anomalous magnetic moment from a massless neutral
mediator is equal for both electrons and muons.

In other cases, the one-loop new physics contributions to $a_e$ can be
obtained from $a_\mu$ simply by replacing the mass of the muon with
the mass of the electron. In the models we consider, this amounts to
multiplying the 
result by $\displaystyle \frac{m_{e}^2}{m_{\mu}^2}\approx 2\times 
10^{-5}$. 
Note that the current uncertainty in $a_e$ is $\delta 
a_e = 0.8\times 10^{-12}$~\cite{PDG2012, Hanneke:2008tm}, where the dominant
uncertainty arises from the measurement of $\alpha$, the fine
structure constant. 
The relative sensitivity of
the two measurements can then be seen as
\begin{align}
  \frac{
  {\hat{a}_e}/ {\delta_e}
  }
  {
  {\hat{a}_\mu}/{\delta_\mu}
  }
  &\sim
  \frac{\delta_\mu}{\delta_e}
  \times
  \frac{m_e^2}
  {m_\mu^2}
  \sim
  0.02
  \ .
\end{align}
We see that even though it is a more precise measurement, 
at present $a_{e}$ is not as sensitive
a probe of new physics as $a_{\mu}$, except for very light mediators.

\subsection{Charged Mediator Diagrams}
\label{ssec:cmd}
\begin{figure}[th]
\centering
\vspace{10mm}
\subfloat[]{
\centering
\begin{fmffile}{vecchrg}
\begin{fmfgraph*}(95,85)\fmfkeep{vecchrg}
\fmfpen{0.7thick}
\fmfstraight\fmftop{p1,g,p2}\fmfcurved\fmfbottom{i1,o1}
\fmfv{l=$\mu$}{i1}\fmfv{l=$\mu$}{o1}\fmfv{l=$\gamma$}{g}
\fmf{fermion,tension=0.8}{i1,v1}\fmf{fermion,label=$F$,label.side=left,tension=0.2}{v1,v2}
\fmf{fermion,tension=0.8}{v2,o1}
\fmf{photon,tension=0.7}{g,v3}
\fmf{wiggly,left=0.5,label=$V$,label.side=left,tension=0.2}{v1,v3,v2}
\fmfv{decor.shape=circle,decor.filled=full,decor.size=1.7thick}{v1}
\fmfv{decor.shape=circle,decor.filled=full,decor.size=1.7thick}{v2}
\fmfv{decor.shape=circle,decor.filled=full,decor.size=1.7thick}{v3}
\end{fmfgraph*}
\end{fmffile}
\label{fig:vecchrg}
}
\qquad\qquad\qquad \qquad
\subfloat[]{
\centering
\begin{fmffile}{vecneut}
\begin{fmfgraph*}(95,85)\fmfkeep{vecneut}
\fmfpen{0.7thick}
\fmfstraight\fmftop{p1,g,p2}\fmfcurved\fmfbottom{i1,o1}
\fmfv{l=$\mu$}{i1}\fmfv{l=$\mu$}{o1}\fmfv{l=$\gamma$}{g}
\fmf{fermion,tension=0.8}{i1,v1}\fmf{wiggly,label=$V$,label.side=left,tension=0.2}{v1,v2}
\fmf{fermion,tension=0.8}{v2,o1}
\fmf{photon,tension=0.7}{g,v3}
\fmf{fermion,left=0.5,label=$F$,label.side=left,tension=0.2}{v1,v3,v2}
\fmfv{decor.shape=circle,decor.filled=full,decor.size=1.7thick}{v1}
\fmfv{decor.shape=circle,decor.filled=full,decor.size=1.7thick}{v2}
\fmfv{decor.shape=circle,decor.filled=full,decor.size=1.7thick}{v3}
\end{fmfgraph*}
\end{fmffile}
\label{fig:vecneut}
}
\caption{One loop contributions to $a_{\mu}$ involving a massive
vector $V^{\mu}$ and massive fermion $F$. In (a) the DM is fermionic
while in (b) it is a vector. }\label{fig:vec}
\end{figure}
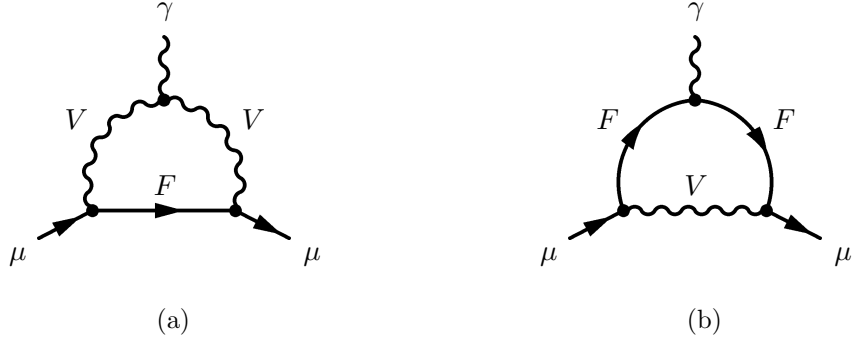

We begin by considering the case of DM-lepton scattering through a charged mediator, as shown in Fig. \ref{fig:tchan}. This scenario encompasses several different theories. In this section, it is convenient to group the different theories by the form of the interaction rather than by the type of DM. For instance, the term
\begin{align}
  \mathcal{L}_{\text{vector}}
  &=
  \bar{\mu}\gamma^{\nu}\left(a+ b\gamma^5 \right)FV_{\nu} 
  +\text{h.c.}\label{eq:vecint}
 \end{align}
describes a renormalizable interaction between the muon, a massive 
fermion $F$ and a massive vector $V^{\nu}$. Depending on the model, 
either $F$ or $V^{\mu}$ constitutes the DM. The leading 
contribution to $a_{\mu}$ from interactions of this form arise 
from Fig. \ref{fig:vecchrg} in the case of fermionic DM, or from Fig. 
\ref{fig:vecneut} for vector DM.

To keep track of each diagram's contribution to $a_{\mu}$ the labeling 
$\hat{a}_{\mu}^{\text{Med,DM}}$ will be used for each of the charged mediator 
contributions. The first term in the superscript refers to the spin of 
the  mediator while the second refers to the spin 
of the DM. For instance the diagram in Fig. \ref{fig:vecchrg} is the 
dominant contribution to $\hat{a}_{\mu}^{V,F}$, corresponding to the case of a 
fermionic DM candidate whose scattering off leptons is mediated by a 
massive vector.

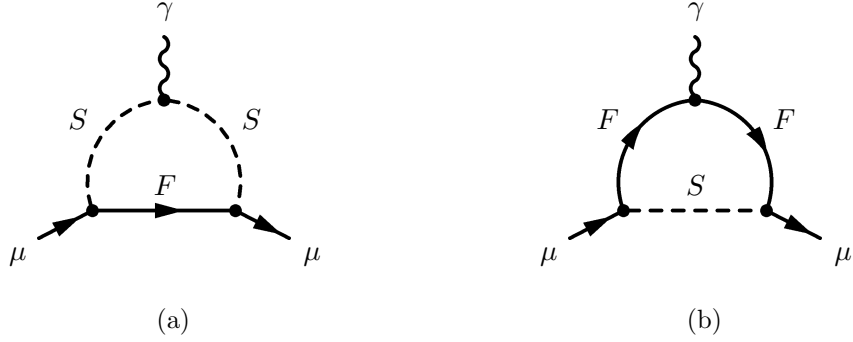
\begin{figure}[th]
\centering
\vspace{10mm}
\subfloat[ ]{
\centering
\begin{fmffile}{scalchrg}
\begin{fmfgraph*}(95,85)\fmfkeep{scalchrg}
\fmfpen{0.7thick}
\fmfstraight\fmftop{p1,g,p2}\fmfcurved\fmfbottom{i1,o1}
\fmfv{l=$\mu$}{i1}\fmfv{l=$\mu$}{o1}\fmfv{l=$\gamma$}{g}
\fmf{fermion,tension=0.8}{i1,v1}\fmf{fermion,label=$F$,label.side=left,tension=0.2}{v1,v2}
\fmf{fermion,tension=0.8}{v2,o1}
\fmf{photon,tension=0.7}{g,v3}
\fmf{dashes,left=0.5,label=$S$,label.side=left,tension=0.2}{v1,v3,v2}
\fmfv{decor.shape=circle,decor.filled=full,decor.size=1.7thick}{v1}
\fmfv{decor.shape=circle,decor.filled=full,decor.size=1.7thick}{v2}
\fmfv{decor.shape=circle,decor.filled=full,decor.size=1.7thick}{v3}
\end{fmfgraph*}
\end{fmffile}
\label{fig:scalchrg}
}
\qquad\qquad\qquad\qquad
\subfloat[ ]{
\centering
\begin{fmffile}{scalneut}
\begin{fmfgraph*}(95,85)\fmfkeep{scalneut}
\fmfpen{0.7thick}
\fmfstraight\fmftop{p1,g,p2}\fmfcurved\fmfbottom{i1,o1}
\fmfv{l=$\mu$}{i1}\fmfv{l=$\mu$}{o1}\fmfv{l=$\gamma$}{g}
\fmf{fermion,tension=0.8}{i1,v1}\fmf{dashes,label=$S$,label.side=left,tension=0.2}{v1,v2}
\fmf{fermion,tension=0.8}{v2,o1}
\fmf{photon,tension=0.7}{g,v3}
\fmf{fermion,left=0.5,label=$F$,label.side=left,tension=0.2}{v1,v3,v2}
\fmfv{decor.shape=circle,decor.filled=full,decor.size=1.7thick}{v1}
\fmfv{decor.shape=circle,decor.filled=full,decor.size=1.7thick}{v2}
\fmfv{decor.shape=circle,decor.filled=full,decor.size=1.7thick}{v3}
\end{fmfgraph*}
\end{fmffile}
\label{fig:scalneut}
}
\caption{One loop contributions to $a_{\mu}$ involving a massive
scalar $S$ and massive fermion $F$. In (a) the DM is fermionic while
in (b) it is a scalar. }\label{fig:scal}
\end{figure}

The other charged mediator interaction is of the form
\begin{align}
\mathcal{L}_{\text{scalar}}=\bar{\mu}\left(a+b\gamma^5 \right)FS +\text{h.c.}\label{eq:scalint}
 \end{align}
where $S$ denotes a massive scalar. This leads to the diagram shown in 
Fig. \ref{fig:scalchrg} if the DM is a fermion, or to the diagram shown in 
\ref{fig:scalneut} if the DM is a scalar.

In the diagrams considered in this subsection the mediating particle is necessarily electrically charged. Therefore, if the mediator is a fermion it must be Dirac, while if it is a scalar or a vector, it must be complex. No such restriction applies to the DM particle. Therefore, if the DM is 
a fermion it could be either Majorana or Dirac, while if it is a boson 
it could either real or complex.

The contributions to $a_{\mu}$ from these diagrams have been calculated 
exactly in \cite{Leveille1978} and \cite{Moore1985}. However, as the 
mass of the DM $m_{\text{DM}}$ and the mediator $m_{\text{Med}}$ are 
expected to be much larger than the mass of the muon $m_{\mu}$, the 
leading order effects can be most clearly seen by expanding the results 
in powers of the small parameter $\varepsilon\equiv 
m_{\mu}/m_{\text{Med}}$. A shorthand designating the ratio $r\equiv 
m_{\text{DM}}/m_{\text{Med}}$ will also be used.

We first give the contributions involving the massive vector. From Fig. 
\ref{fig:vecchrg} we obtain
\begin{align}
  \hat{a}_{\mu}^{V,F}
  =
\frac{\varepsilon}{16\pi^2}&
  \left\{\frac{\varepsilon(|a|^2+|b|^2)} 
  {3(1-r^2)^4}
  \left[10-43r^{2}+78r^4-49r^6+4r^8 +18r^6\ln(r^2)\right]\right. 
  \nonumber\\&\;\left.
  -\frac{r(|a|^2-|b|^2)}{(1-r^2)^3}
  \left[ 4-15r^2+12r^4-r^6- 6r^4\ln(r^2)\right]\right\}
  +\mathcal{O}(\varepsilon^3)
  \label{eq:avf}
\end{align}
while from Fig. \ref{fig:vecneut} we find
\begin{align}
\hat{a}_{\mu}^{F,V}=
-\frac{\varepsilon}{16\pi^2r^2}
&\left\{ \frac{\varepsilon(|a|^2+|b|^2)}{3(1-r^2)^4}\left[5-14r^2+39r^4-38r^6 +8r^8+18r^4\ln(r^2)\right]\right.\nonumber\\
&\left.\;-\frac{|a|^2-|b|^2}{(1-r^2)^3}\left[1+3r^4-4r^6+6r^4\ln(r^2)\right] \right\}+\mathcal{O}(\varepsilon^3).\label{eq:afv}
\end{align}
The contributions from the scalar interactions in Figures 
\ref{fig:scalchrg} and \ref{fig:scalneut} are respectively
 \begin{align}
\hat{a}_{\mu}^{S,F}
&=
-\frac{\varepsilon}{16\pi^2}
\left\{\frac{\varepsilon(|a|^2+|b|^2)}{3(1-r^2)^4} \left[1-6r^2+3r^4+2r^6-6r^4\ln(r^2)\right]\right. \nonumber\\ &\left.\phantom{-\frac{\varepsilon}{16\pi^2}}\;\;+\frac{r(|a|^2-|b|^2)}{(1-r^2)^3}\left[ 1-r^4+2r^2\ln(r^2)\right]\right\}+\mathcal{O}(\varepsilon^3),\label{eq:asf}\\
\hat{a}_{\mu}^{F,S}&=\frac{\varepsilon}{16\pi^2}\left\{\frac{\varepsilon(|a|^2+|b|^2)}{3(1-r^2)^4} \left[1-6r^2+3r^4+2r^6-6r^4\ln(r^2)\right] \right.\nonumber\\ &\left.\phantom{\frac{\varepsilon}{16\pi^2}}\;\;+\frac{|a|^2-|b|^2}{(1-r^2)^3}\left[1-4r^2+3r^4 -2r^4\ln(r^2)\right]\right\}+\mathcal{O}(\varepsilon^3).\label{eq:afs}
\end{align}

In all of the above relations the $r$ dependent functions contained 
within the square brackets are never negative and hence never change the 
sign of the contribution. Therefore, in each case the sign of the 
correction to $a_{\mu}$ is completely determined by the relative sizes 
of $a$ and $b$. An interesting feature pointed out by \cite{Grifols1982} 
in the context of supersymmetric theories, but which is completely 
general, is that the contribution to $a_{\mu}$ from a coupling to the 
muon that respects chiral symmetry, $a = \pm b$, is suppressed by order 
$\varepsilon$ relative to the results of an interaction that violates this
symmetry.

\subsection{Neutral Mediator Diagrams\label{ssec:nmd}}

We now consider DM-lepton scattering mediated by a neutral particle,
as shown in Fig. \ref{fig:schan}. The interactions between the muon
and a vector or a scalar mediator are of the same form as in
\eqref{eq:vecint} and \eqref{eq:scalint} respectively, but with the
fermion $F$ replaced by a muon $\mu$. Because the contributions to
$a_{\mu}$ in these models are insensitive to the nature of the DM
particle, we simply label the contributions by the spin of the
mediator, i.e. $\hat{a}_{\mu}^{V}$ for a vector mediator.

\begin{figure}[th]
\centering
\vspace{10mm}
\subfloat[ ]{
\centering
\begin{fmffile}{sscalneut}
\begin{fmfgraph*}(95,85)\fmfkeep{sscalneut}
\fmfpen{0.7thick}
\fmfstraight\fmftop{p1,g,p2}\fmfcurved\fmfbottom{i1,o1}
\fmfv{l=$\mu$}{i1}\fmfv{l=$\mu$}{o1}\fmfv{l=$\gamma$}{g}
\fmf{fermion,tension=0.8}{i1,v1}\fmf{dashes,label=$S$,label.side=left,tension=0.2}{v1,v2}
\fmf{fermion,tension=0.8}{v2,o1}
\fmf{photon,tension=0.7}{g,v3}
\fmf{fermion,left=0.5,label=$\mu$,label.side=left,tension=0.2}{v1,v3,v2}
\fmfv{decor.shape=circle,decor.filled=full,decor.size=1.7thick}{v1}
\fmfv{decor.shape=circle,decor.filled=full,decor.size=1.7thick}{v2}
\fmfv{decor.shape=circle,decor.filled=full,decor.size=1.7thick}{v3}
\end{fmfgraph*}
\end{fmffile}
\label{fig:sscalneut}
}
\qquad\qquad\qquad\qquad
\subfloat[ ]{
\centering
\begin{fmffile}{svecneut}
\begin{fmfgraph*}(95,85)\fmfkeep{svecneut}
\fmfpen{0.7thick}
\fmfstraight\fmftop{p1,g,p2}\fmfcurved\fmfbottom{i1,o1}
\fmfv{l=$\mu$}{i1}\fmfv{l=$\mu$}{o1}\fmfv{l=$\gamma$}{g}
\fmf{fermion,tension=0.8}{i1,v1}\fmf{wiggly,label=$V$,label.side=left,tension=0.2}{v1,v2}
\fmf{fermion,tension=0.8}{v2,o1}
\fmf{photon,tension=0.7}{g,v3}
\fmf{fermion,left=0.5,label=$\mu$,label.side=left,tension=0.2}{v1,v3,v2}
\fmfv{decor.shape=circle,decor.filled=full,decor.size=1.7thick}{v1}
\fmfv{decor.shape=circle,decor.filled=full,decor.size=1.7thick}{v2}
\fmfv{decor.shape=circle,decor.filled=full,decor.size=1.7thick}{v3}
\end{fmfgraph*}
\end{fmffile}
\label{fig:svecneut}
}
\caption{One loop contributions from models with neutral mediators to $a_{\mu}$ by (a) a scalar $S$ and (b) a vector $V^{\mu}$. }\label{fig:sloops}
\end{figure}
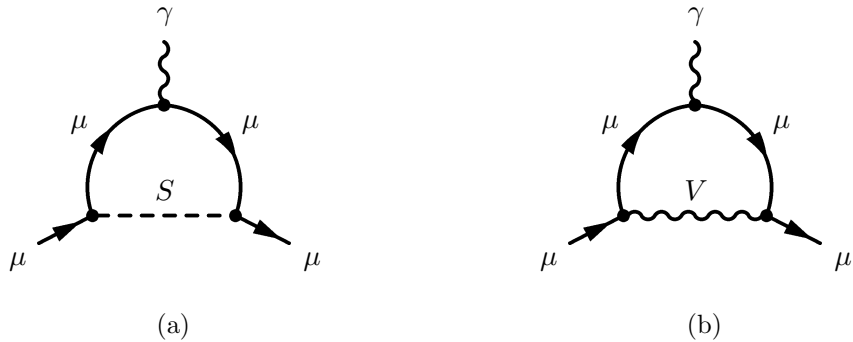

The interactions of the neutral mediator models are nearly
identical to the charged mediator models, and in principle the exact expressions
from the calculation above can be adapted for this case.
However, the expansion in terms of the ratio of the muon mass to the
charged particle running in the loop, in this case the muon
itself, is obviously not valid here. We continue using $\varepsilon$
to denote the ratio of masses of the muon and the neutral mediator,
and present the leading order result in this parameter,
\begin{align}
  \hat{a}_{\mu}^{V}
  &=
  \frac{\varepsilon^2}{4\pi^2}
  \left[
  -\frac{2}{3}(|a|^2+|b|^2)
  + (|a|^2-|b|^2) 
  \right]
  +\mathcal{O}(\varepsilon^3),
  \label{eq:av}
  \\
  \hat{a}_{\mu}^{S}
  &=
  \frac{\varepsilon^2}{8\pi^2}
  \left[
  \frac{1}{3}(|a|^2+|b|^2)- 
  \left( \frac{3}{2}+\ln\varepsilon^2 \right) 
  (|a|^2-|b|^2) 
  \right]
  +\mathcal{O}(\varepsilon^3).
  \label{eq:as}
\end{align}
Notice that the contribution from interactions that respect chiral 
symmetry is no longer suppressed by an additional power of $\varepsilon$ 
relative to the contribution from couplings that violate this symmetry. In 
the case of a scalar mediator, the chiral symmetry violating effects are 
only logarithmically enhanced, while in the case of a vector mediator they 
are of the same order. When the mediator is lighter than or comparable in 
mass to the muon, the small $\varepsilon$ approximation does not apply, and 
we use the full expressions for our numerical analysis of constraints.

The muon $g-2$ is sensitive only to couplings of the DM and mediators
to the muon. However, other constraints depend on the flavor structure
of the couplings to all leptons. Therefore, to compare the muon $g-2$
with these limits, we must make some assumptions about the flavor
structure of the models under consideration. The non-observation of
flavor-changing processes in charged leptons puts strong constraints
on the couplings of DM to leptons. We assume that each lepton couples
to the DM candidate flavor diagonally and with universal strength.
Accordingly, we assume that there is a separate charged mediator
corresponding to each lepton flavor, and that there is no mixing
between different flavors of mediators.  Another possibility that also
satisfies flavor bounds, but which we will not consider in this paper,
is to have a separate DM species corresponding to each flavor of
lepton, but only a single mediator~\cite{Agrawal:2011ze}.

\section{Limits from Direct Detection\label{sec:compbnds}}

In this section we determine the limits on leptophilic DM from direct
detection experiments. In these theories scattering off nuclei only
arises at loop level. Therefore, we expect that the direct detection
bounds will most significantly constrain theories where DM-nucleon
scattering is spin-independent, and arises at one loop. In our
analysis we use the recently released LUX
results~\cite{Akerib:2013tjd} to generate the bounds.

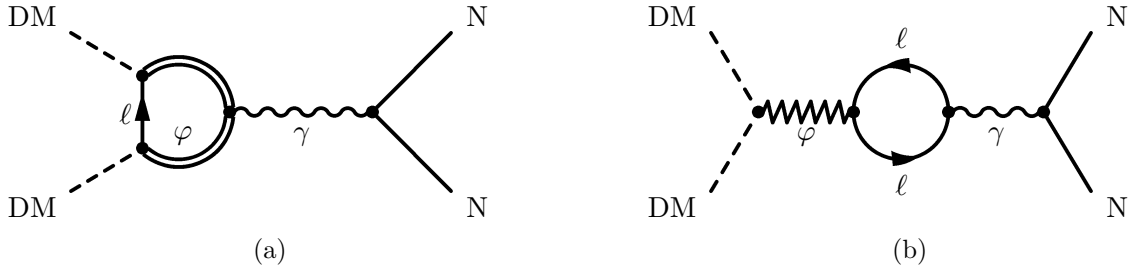
\begin{figure}[tp]
\centering
\vspace{10mm}
\subfloat[ ]{
\centering
\begin{fmffile}{ddtchan}
\begin{fmfgraph*}(180,60)\fmfkeep{ddtchan}
\fmfpen{0.7thick}
\fmfleft{i1,i2}\fmfright{o1,o2}
\fmfv{l=DM}{i1}\fmfv{l=DM}{i2}\fmfv{l=N}{o1}\fmfv{l=N}{o2}
\fmf{dashes,tension=1.1}{i1,v1}\fmf{dashes,tension=1.1}{i2,v2}
\fmf{fermion,tension=0.2}{v1,v2}
\fmf{double,right=0.67,label=$\varphi$,label.side=left,tension=0.9}{v1,v3}
\fmf{double,left=0.67,tension=0.9}{v2,v3}
\fmf{photon,label=$\gamma$,tension=1.1}{v3,v4}
\fmf{plain,tension=1}{v4,o1}\fmf{plain,tension=1}{v4,o2}
\fmfvn{decor.shape=circle,decor.filled=full,decor.size=1.7thick}{v}{4}
\fmfv{l=$\ell$,l.d=10,l.a=115}{v1}
\end{fmfgraph*}
\end{fmffile}
\label{fig:ddtchan}
}
\qquad\qquad
\subfloat[ ]{
\centering
\begin{fmffile}{ddschan}
\begin{fmfgraph*}(180,60)\fmfkeep{ddschan}
\fmfpen{0.7thick}
\fmfleft{i1,i2}\fmfright{o1,o2}
\fmfv{l=DM}{i1}\fmfv{l=DM}{i2}\fmfv{l=N}{o1}\fmfv{l=N}{o2}
\fmf{dashes}{i1,v1}\fmf{dashes}{i2,v1}
\fmf{zigzag,label=$\varphi$}{v1,v2}
\fmf{fermion, label=$\ell$, right,tension=0.5}{v2,v3,v2}
\fmf{photon,label=$\gamma$}{v3,v4}
\fmf{plain}{v4,o1}\fmf{plain}{v4,o2}
\fmfvn{decor.shape=circle,decor.filled=full,decor.size=1.7thick}{v}{4}
\end{fmfgraph*}
\end{fmffile}
\label{fig:ddschan}
}
\caption{Leading order processes for DM nucleon scattering for (a)
charged and (b) neutral mediators. Note that there is a second diagram
at the same order with the lepton $\ell$ coupling to the photon in the
charged mediator case. }\label{fig:ddchan}
\end{figure}

In Fig. \ref{fig:ddchan} we see the leading contributions to a DM
particle $\chi$ scattering with a nucleus $N$, arising from a photon
exchange. In general, the direct detection cross section receives
contributions from all the SM leptons running in the loop.  As stated
above, we focus on theories where the couplings of DM to the SM
leptons are universal and flavor diagonal.

The velocity of incident DM particles is expected to be small ($v/c\sim 
10^{-3}$), which implies that the momentum transfer in direct detection 
experiments will be of order 10-50 MeV. Since the momentum transfer in 
this process is so small, it is convenient to work in an effective 
theory with the charged particles in the loop that mediate scattering 
integrated out. The effective theory will be consistent provided that 
the momentum transfer in the scattering process is smaller than the mass 
of the particles in the loop. Although this condition is not strictly 
satisfied in the case of the electron, it is straightforward to correct 
for this. In the case of charged mediators, in the effective theory the 
DM candidate couples directly to the photon through an effective vertex.

Similar considerations apply to the neutral mediator case, provided the
mass of the neutral mediator is more than the typical momentum
transfer. Of course, since DM-nucleon scattering arises via mixing of
the mediator with the photon, only the neutral vector contributes.
The mixing is radiatively generated by a loop diagram that involves
particles charged under both the photon and the new vector boson. At
low energies, the only particles in the loop that contribute to mixing
are the charged leptons of the SM. The loop diagram is logarithmically
divergent, and needs to be regulated. For concreteness, we assume that
the diagram is cut off at a scale $\Lambda_V$, which we take to be of
order the weak scale. After integrating out the mediator and the
charged leptons, the DM again couples to the photon
through an effective vertex at low energies.

The case of very light neutral mediator ($m \lesssim 30$ MeV)  is
slightly more subtle. We can no longer integrate out the mediator for
typical scattering at direct detection experiments. In this case, it
is more convenient to eliminate the mixing between the mediator and
the photon by a redefinition of the photon field, which leads to a
direct coupling of the SM quarks to the neutral mediator.  This
translates into a tree-level interaction of the DM with quarks in the
low energy theory, mediated by the new vector boson.

It follows that the direct detection bounds can be translated into
limits on the effective operators that couple DM to the photon, and
then into constraints on the parameters of the theory. These
constraints are expected to be most significant for theories that give
rise to effective operators that generate spin-independent DM-nucleon
scattering at one loop. In what follows we consider the DM candidates
of different spins and identify the relevant effective operators that
lead to spin-independent DM-nucleon interactions. Those theories with
no such operator are not expected to be significantly constrained by
the current limits from direct detection.

The effective operators generated by integrating out the particles in 
the loop respect the symmetries of the underlying theory. Therefore,
we need only consider effective operators that respect these
symmetries. In particular, since we are restricting our discussion to
interactions that are invariant under $CP$, we need only consider
effective operators that respect $CP$.

\subsection{Self-conjugate particles}
It can be shown quite generally that in the case of self-conjugate (real
scalar, Majorana fermion, real vector) leptophilic DM,
the direct detection signals are suppressed. 
A self-conjugate particle couples to a single photon
only through an anapole moment \cite{
Zeldovich,
Nieves:1981zt,
Kayser:1982br,
Radescu:1985wf,
Boudjema:1988zs,
Boudjema:1990st}. In particular, for $CP$-conserving interactions, 
the
scattering amplitude of the DM particles with the quark
electromagnetic current ($\bar{q} \gamma^\mu q$) can be written in the
following form,
\begin{align}
  \mathcal{M}
  \sim 
  \left(k^2 s_\mu - (k\cdot s) k_\mu \right)
  \frac{1}{k^2}
  \bar{q} \gamma^\mu q
  \ .
\end{align}
 Here $p_1$ and $p_2$ are the incoming and outgoing DM momenta 
respectively, and $k = p_2 - p_1$ is the momentum transfer. The quantity 
$s_\mu$ depends on the spins and momenta of the incoming and outgoing DM 
particles. As an example, for a Majorana fermion, $s^\mu = 
\langle\chi(p_2)| \bar{\chi}\gamma^\mu \gamma^5 \chi | \chi(p_1) 
\rangle$. The second term above does not contribute to the scattering 
process (due to Ward identities). The $k^2$ in the first term cancels 
against the photon propagator, giving rise to a contact interaction with 
the quark current. For direct detection, this implies that the 
scattering amplitude with the nucleus is given by
 \begin{align}
  \mathcal{M}
  &\sim
  \sum_q
  \langle N_f |
  Q \bar{q} \gamma^\mu q
  |N_i\rangle
  s_\mu
  \ .
\end{align}
 Since the anapole moment is $P$-odd (a symmetry respected by the 
electromagnetic coupling), it follows that this contact interaction will 
lead to $p$-wave scattering amplitudes that are suppressed in the 
non-relativistic limit. Therefore, we expect that the direct detection 
constraints on self-conjugate DM particles coupling to the nucleons via a 
photon will be relatively weak.

These conclusions may also be directly obtained from a study of the 
possible effective operators. We limit ourselves to operators of
dimension up to six.  It is easily seen that there are no 
operators that couple real scalar DM to a single photon. Such an operator 
would schematically contain two DM field factors $\chi^2$, the photon 
field strength $F_{\mu\nu}$ (or its dual), and derivatives. The 
antisymmetry of $F^{\mu\nu}$, however, makes any operator of this type 
vanish identically.

As is well-known, the bilinears $\bar{\chi}\gamma^{\mu}\chi$ and
$\bar{\chi}\sigma_{\mu\nu}\chi$ both vanish identically for Majorana
fermions.  The only surviving bilinear is proportional to
$\bar{\chi}\gamma^{\mu}\gamma^5\chi$ which gives velocity suppressed
matrix elements with the electromagnetic current of the nucleus.

Finally for  real vector DM the 
dimension four operator $\chi_{\mu}\chi_{\nu}F^{\mu\nu}$ vanishes by the 
antisymmetry of the Maxwell tensor. We can apply two derivatives as 
in the real scalar case, but find that the resulting operator 
violates $CP$. This leads us to consider the operator
\begin{align}
   \mathcal{O}_V
   &=
   \partial_{\mu}\chi^{\alpha}\partial_{\alpha}\chi_{\nu} 
   F_{\sigma\rho}\varepsilon^{\mu\nu\sigma\rho}.
\end{align} 
Note that any other choice of contraction of indices vanishes either
due to the antisymmetry of $\varepsilon^{\mu\nu\sigma\rho}$ or the
on-shell constraint $\partial_{\alpha}\chi^{\alpha}=0$. The resulting
amplitude is
\begin{align}
  \mathcal{M}
  &\sim
  \sum_{q}
  \langle N_f|Q\bar{q}\gamma_{\rho}q|N_{i}\rangle 
  \frac{1}{k^2}
  \varepsilon^{\mu\nu\sigma\rho}
  k_{\sigma}(p_2+p_1)_{\mu}k_{\alpha}
  \left[  \epsilon^*_{\nu}(p_2)\epsilon^{\alpha}(p_1) 
  -\epsilon_{\nu}(p_1)\epsilon^{*\alpha}(p_2)
  \right]
  \label{eq:vddamp}
\end{align}
 where the $\epsilon_{\nu}(p)$ are polarization vectors of $\chi_{\nu}$. 
The $k_\sigma k_\alpha$ suppression is overcome by the photon propagator. 
However, the epsilon-tensor contraction of $(p_1 + p_2)$ with $\bar{q} 
\gamma^\mu q$ vanishes in the limit that the velocities of the incoming 
particles tend to zero. Therefore, this amplitude is suppressed in the 
non-relativistic limit.

We conclude that leptophilic DM particles which are real scalars, Majorana 
fermions or real vectors are not generally expected to be significantly constrained 
by direct detection searches. The DM candidates that give rise to sizable 
spin-independent cross sections in the non-relativistic limit are complex 
scalars and Dirac fermions. We study those cases in more detail.

\subsection{Complex scalar DM\label{sssec:ddSDM}}
If the DM is a complex scalar there is a non-vanishing effective operator, 
namely
 \begin{align}
\mathcal{O}_{S}&=A_Si\left[\partial_{\mu}\chi^{\ast}\partial_{\nu}\chi- \partial_{\nu}\chi^{\ast}\partial_{\mu}\chi \right]F^{\mu\nu}.
 \end{align}
Note that if we replace $F^{\mu\nu}$ by 
$F_{\alpha\beta}\varepsilon^{\alpha\beta\mu\nu}$ the corresponding 
operator is odd under $CP$. From $\mathcal{O}_S$ we find the complex 
scalar DM contribution to the direct detection amplitude,
\begin{align}
  \mathcal{M}_{\mathcal{O}_S}
  &=
  -A_S e
  \sum_q\langle N_f|Q\bar{q}\gamma^{\nu}q|N_i\rangle (p_2+p_1)_{\nu}
  \equiv
  \widetilde{\lambda}
  \sum_q\langle N_f|Q\bar{q}\gamma^{\nu}q|N_i\rangle (p_2+p_1)_{\nu}
  \label{eq:ddscaleff}
\end{align} 
where the sum over $q$ denotes a sum over quark bilinears, $p_1$ and
$p_2$ the DM momenta before and after scattering, and $Q$ is the
charge each quark in units of $e$.

\subsection{Dirac fermion DM\label{sssec:ddfrmDM}}
For a Dirac fermion DM, the lowest dimension operator we can write 
down which preserves $CP$ and gauge invariance is
\begin{align}
  \mathcal{O}
  &=iA_5\,\bar{\chi}\sigma_{\mu\nu}\chi F^{\mu\nu}.\label{eq:dddim5}
\end{align} 
However, this effective operator violates the chiral symmetry 
associated with $\chi$, and will not be generated in theories where the 
couplings of DM respect this symmetry. When present this operator leads 
to the amplitude
\begin{align}
  \mathcal{M}
  &=2A_5 e\sum_q\langle N_f|Q\bar{q}\gamma^{\nu}q|N_i\rangle
  \frac{k^{\mu}}{k^2}\bar{u}(p_2)\sigma_{\mu\nu}u(p_1)\label{eq:dipole}
\end{align}
 where $k=p_2-p_1$ is the momentum transfer. This amplitude gives rise 
to dipole-charge and dipole-dipole interactions between the DM and the 
nucleus \cite{Barger201174,Chang:2010en}. The dipole-charge interaction 
is enhanced by the atomic number $Z$ for heavy nuclei, and hence 
dominates. It is important to note that the recoil spectrum arising from 
this interaction is distinct from that of the familiar contact 
interaction, and simply scaling the WIMP-nucleon cross section limit 
does not give accurate results in this case.

In the simplified models we consider, the dominant contribution arises at
dimension six. At this order, we have two possible $CP$-even operators,
\begin{align}
  \mathcal{O}_1
  &=
  \left[\,\bar{\chi}\gamma^{\mu}(c+d\gamma^{5})\partial^{\nu}\chi 
  +\text{h.c.} \right]F_{\mu\nu} ,\\
  \mathcal{O}_2
  &=
  \left[i\bar{\chi}\gamma^{\mu}(c+d\gamma^{5})\partial^{\nu}\chi 
  +\text{h.c.}  \right]F^{\sigma\rho}\varepsilon_{\mu\nu\sigma\rho} 
\end{align}
where the factors of $i$ have been chosen to make the effective theory
coefficients $c$ and $d$ real. These operators were considered in
Appendix A of \cite{Agrawal:2011ze} for the case of charged scalar
mediator. 

The above operators lead to the amplitudes 
\begin{align}
  \mathcal{M}_{\mathcal{O}_1}
  &=
  -e c \sum_{q}\langle N_f|Q\bar{q}\gamma^{\mu}q|N_{i}\rangle 
  \bar{u}(p_2)\gamma_{\mu}u(p_1),\label{eq:ddtdampo1}
  \\
  \mathcal{M}_{\mathcal{O}_2}
  &= 4ied\sum_{q}
  \langle N_f|Q\bar{q}\gamma_{\alpha}q|N_{i}\rangle 
  \frac{1}{k^2}\bar{u}(p_2)
  \left[m_{\chi}\sigma^{\alpha\beta}k_{\beta} 
  +\frac{i}{2}k^2\gamma^{\alpha}\right]u(p_1).
  \label{eq:ddtdampo2}
\end{align}
where we have neglected velocity suppressed terms.
Notice that $\mathcal{O}_2$ leads to interactions of 
the charge-charge, dipole-charge and dipole-dipole form.
We use the charge-charge interaction to
set limits. The generic amplitude for this can be parametrized as
\begin{align}
  \mathcal{M}
  &=
  \widetilde{\lambda}
  \,
  \sum_q
  \langle N_f|Q\bar{q}\gamma_{\alpha}q|N_{i}\rangle 
  \,\bar{u}(p_2) \gamma^\alpha u(p_1)
  \ .
  \label{eq:dddiraceff}
\end{align}

Since the argument of logarithm is large, a leading-log approximation
(as employed in \cite{Kopp:2009et}) suffices to place limits on these
models.

\subsection{Neutral vector mediator}
We take a brief detour to consider the case of the neutral vector
mediator separately, since
it has a few distinct qualitative features. We choose
the case of Dirac fermion DM for illustration. To begin, we assume that
both the mediator and the leptons are heavier than the typical
momentum transfer at direct detection experiments. This is obviously
not correct for the electron and very light mediators, and we will
subsequently correct for this.

%We start with writing
%the Lagrangian at a scale $\Lambda_V$, the scale at which additional
%physics associated with the neutral vector, such as
%additional fermions charged under it, are expected to appear. 
%The relevant terms in
%the Lagrangian at this scale are,
%\begin{align}
%  \mathcal{L}
%  &=
%  \lambda \bar{\chi}\gamma^\mu \chi V_\mu 
%  +\lambda \bar{\ell}\gamma^\mu \ell V_\mu
%  - \frac12 m_V^2 V^\mu V_\mu
%  -\frac14 F_{\mu \nu} F^{\mu \nu}
%  -\frac14 F'_{\mu \nu} F'^{\mu \nu}
%  +e \bar{\ell}\gamma^\mu \ell A_\mu \ .
%\end{align}
%where $F_{\mu\nu}$ and $F'_{\mu\nu}$ are the field strength tensors
%for the photon and the neutral mediator respectively.

We first consider the case where the mediator is heavier than all the leptons.
Just above the scale of the mediator mass, the relevant terms in the
effective Lagrangian are given by,
\begin{align}
  \mathcal{L}_{eff}
  &=
  \lambda \bar{\chi}\gamma^\mu \chi V_\mu 
  +\lambda \bar{\ell}\gamma^\mu \ell V_\mu
  -\frac12 m_V^2 V^\mu V_\mu
  +e \bar{\ell}\gamma^\mu \ell A_\mu 
  -\frac14 F_{\mu \nu} F^{\mu \nu}
  -\frac14 F'_{\mu \nu} F'^{\mu \nu}
  + \tilde{\epsilon} F'_{\mu \nu} F^{\mu \nu} \ .
\end{align}
Since the leptons are charged under both the $A_\mu$ and $V_\mu$, they
contribute to the kinetic mixing between the photon and the neutral mediator
at one loop. To set conservative limits, we assume that this
constitutes the dominant contribution to $\tilde{\epsilon}$, so that
$\tilde{\epsilon}$ in the leading log approximation is given by,
\begin{align}
  \tilde{\epsilon}
  &=
  \frac{\lambda e}{24\pi^2}
  \sum_l\log \frac{m_{V}^2}{\Lambda_V^2} \ .
\end{align}
For concreteness, we take the cut-off $\Lambda_V$ to be $1 $ TeV. 

The mediator can be integrated out using its equation of motion.
\begin{align}
(\partial^2 g^{\mu \nu} -\partial^\mu \partial^\nu + m_V^2 g^{\mu\nu})
V_\mu
&=
2 \tilde{\epsilon} \partial_\mu  F^{\mu \nu}
 +
  \lambda \bar{\chi}\gamma^\mu \chi 
  +\lambda \bar{\ell}\gamma^\mu \ell 
\end{align}
Thus, below the mediator mass, the relevant terms at leading order are
\begin{align}
  \mathcal{L}_{eff}
  &=
  \frac{\lambda^2}{m_V^2} \bar{\chi}\gamma^\mu  \chi \,
  \bar{\ell}\gamma^\mu \ell 
  +e\, \bar{\ell}\gamma^\mu \ell A_\mu
  +
  \frac{2 \tilde{\epsilon} \lambda}{m_V^2} \bar{\chi}\gamma^\nu  \chi
  \partial_\mu F^{\mu\nu}
  -\frac14 F_{\mu\nu}F^{\mu\nu}
\end{align}
With this Lagrangian, the DM couples to the photon through
the higher dimensional operator directly, but also via a loop of
leptons. The second contribution is logarithmically divergent, which
is to say that the coupling of DM to photons continues to
run. Below the mass of the leptons, we are only left with,
\begin{align}
  \mathcal{L}_{eff}
  &=
  \frac{2 \epsilon \lambda}{m_V^2} \bar{\chi}\gamma^\nu  \chi
  \partial_\mu F^{\mu\nu}
  -\frac14 F_{\mu\nu}F^{\mu\nu}
\end{align}
The coefficient $\epsilon$ includes logarithms from running between
$\Lambda$ to $m_V$, as well as $m_V$ to $m_l$. In fact, from the full
theory calculation these
terms are easily seen to combine into a single logarithm of the ratio
of scales
$m_{\ell}/\Lambda_V$.
\begin{align}
  \epsilon
  &=
  \frac{\lambda e}{24\pi^2}
  \sum_l\log \frac{m_{\ell}^2}{\Lambda_V^2}
  \label{eq:epsdef}
\end{align}

If the mediator is lighter than the leptons, then the effective
Lagrangian below the lepton masses contains a kinetic-mixing of the
mediator with the photon.
\begin{align}
  \mathcal{L}_{eff}
  &=
  \lambda \bar{\chi}\gamma^\mu \chi V_\mu 
  -\frac12 m_V^2 V^\mu V_\mu
  -\frac14 F_{\mu \nu} F^{\mu \nu}
  -\frac14 F'_{\mu \nu} F'^{\mu \nu}
  + \epsilon F'_{\mu \nu} F^{\mu \nu}
  \label{eq:heavylep}
\end{align}
where $\epsilon$ is defined in Eq.~\eqref{eq:epsdef}.
Below this scale, the DM coupling to photons does not run.
One the mediator is integrated out, we again generate the same
operator as above.
\begin{align}
  \mathcal{L}_{eff}
  &=
  \frac{2 \epsilon \lambda}{m_V^2} \bar{\chi}\gamma^\nu  \chi
  \partial_\mu F^{\mu\nu}
  -\frac14 F_{\mu\nu}F^{\mu\nu}
\end{align}
Thus, we see that the effective operators generated in the neutral
mediator case have the same form as in the charged mediator case, and hence
the results
from that analysis can be applied here as well.

We finally address the assumption that the leptons and the mediator
masses are heavier than the momentum transfer in direct detection
experiments. At LUX, the momentum transfer corresponding to the energy
thresholds is around $|q|\sim30$ MeV. Therefore, the logarithms
appearing in the expression for $\epsilon$, Eq.~\eqref{eq:epsdef}, are
cut off by $|q|$ in the case of the electron,
rather than by the electron mass. 

If the neutral mediator mass is below the typical scattering energies,
$|q|\sim 30$ MeV, it cannot be integrated out.
However, we can eliminate the kinetic
mixing in Eq.~\eqref{eq:heavylep} via a field redefinition $A_\mu \to
A_\mu + 2 \epsilon V_\mu$. After this redefinition, quarks pick up
couplings to the vector mediator, via which DM can scatter off nuclei
at direct detection experiments. Since the mediator is very light, the
propagator $1/(q^2-m^2)$ is dominated by $1/q^2$, and hence the
scattering cross section scales as $1/q^4$.  For setting limits in
this regime, we assume that the scattering is dominated by $|q| \sim
30$ MeV. This level of approximation will suffice for our purposes.

\subsection{Limits}

The differential cross section for charge-charge interactions takes the form
 \begin{align}
\frac{d\sigma}{d \vec{k}^2}=\frac{Z^2\widetilde{\lambda}^2}{4\pi v^2}F^2(\vec{k}^2)\label{eq:diffxsec}
\end{align}
where $Z$ is the charge number of the nucleus and $F(\vec{k}^2)$ is
the charge form factor of the nucleus. The quantity
$\widetilde{\lambda}$ is the coefficient of the charge-charge
interaction
operators in
Eqs. \eqref{eq:ddscaleff} and \eqref{eq:dddiraceff}. This expression
applies to both cases above.

In order to make connection with experimental limits, we evaluate the
zero momentum transfer cross section, which is 
obtained by integrating \eqref{eq:diffxsec} evaluated at $\vec{k}=0$.
\begin{align}
  \sigma^{0}_N
  &=\frac{\mu_N^2Z^2}{\pi}\widetilde{\lambda}^2 \label{eq:ddnucleus}
\end{align}
where $\mu_N$ is the reduced mass of the DM-nucleus system. Further,
since limits are reported in terms of the DM-nucleon cross section, we
rescale the DM-nucleus cross section,
\begin{align}
  \sigma_n^0
  &=
  \sigma_N^0\frac{\mu_n^2}{\mu_N^2}\frac{1}{A^2} 
  =\frac{\mu_n^2Z^2}{\pi A^2}\widetilde{\lambda}^2
  \label{eq:ddnucleon}
\end{align}
where $\mu_n$ is the reduced mass of the DM nucleon system and $A$ is
the number of nucleons in the nucleus.  Thus, given an effective
Lagrangian, limits can be placed on $\widetilde{\lambda}$ using
\eqref{eq:ddnucleon}. In
section \ref{sec:models}, we calculate the coefficient
$\widetilde{\lambda}$ in the context of a set of simplified models,
and use the equation above to
place limits on the parameter space.

\section{Collider constraints}
\label{sec:LEP}
In this section we study collider constraints on leptophilic dark
matter models. 
In our analysis, we assume that 
the DM candidate couples with equal strength to all three flavors of
leptons.  In this scenario, DM particles can be pair-produced
at LEP, in association with hard initial state radiation and
significant missing transverse energy. This is called the monophoton signal. 
Existing analyses focus on
the case of fermionic DM~\cite{Fox:2011fx}. 

Neutral mediators can be resonantly produced at LEP and other $e^+
e^-$ colliders if they are light enough. If they are heavier than 208
GeV, the LEP energy threshold, the LEP limits on four-lepton
operators apply, the so-called ``compositeness bounds"~\cite{PDG2012}.
Charged mediators can be pair-produced at colliders. Studies of direct
slepton pair-production can be translated into limits on these models.
In general, these constraints on parameter space are comparable to,
and often stronger than, the restrictions from $a_{\mu}$.

%We must make one other assumption to compare with the LEP analysis. That 
%work gives bounds in terms of $\sqrt{g_e g_{\text{DM}}}/M$ where $g_e$ 
%and $g_{\text{DM}}$ are the coupling of the mediator to the electron and 
%DM respectively and $M$ is the mass of the mediator. As a simple first 
%guess we assume that the couplings are equal $\displaystyle 
%g_2=g_{\text{DM}}=\frac{\lambda}{2}$.

\subsection{Neutral mediators}
\subsubsection{Compositeness bounds from LEP}
Models with neutral mediators heavier than 208 GeV are constrained by
severe limits on four-lepton contact operators, the compositeness 
bounds~\cite{PDG2012,Compositeness1983}. 
These operators are generated at tree-level when the neutral 
mediator that couples the DM to leptons is integrated out. The bounds are expressed as 
limits on the new physics scale $\Lambda$ that appears as the
coefficient suppressing these higher-dimensional operators.
The operators can be parametrized as~\cite{Kroha92}
\begin{align}
  \mathcal{L}
  &=
  \frac{4\pi}{(1+\delta) \Lambda^2}
  \left[
  \eta_{LL}   \bar{e}_L\gamma^{\mu} e_L\bar{\ell}_L\gamma_{\mu}\ell_L
  + \eta_{RR} \bar{e}_R\gamma^{\mu} e_R\bar{\ell}_R\gamma_{\mu}\ell_R 
  + \eta_{LR} \bar{e}_L\gamma^{\mu} e_L\bar{\ell}_R\gamma_{\mu}\ell_R
  + \eta_{RL} \bar{e}_R\gamma^{\mu} e_R\bar{\ell}_L\gamma_{\mu}\ell_L 
  \right]
  \label{eq:compeq}
\end{align} 
where coefficients $\eta$ characterize the relative contributions of
different chiralities. The factor $\delta$ is 1 when $\ell=e$ and 0 otherwise.

\begin{figure*}[h]
  \vspace{5mm}
  \begin{tabular}{ccc}
    %\displaystyle\begin{adjustbox}{max size={40mm}{20mm},raise= {-1mm}{\height}}
    \begin{tabular}{c}
      \begin{fmffile}{comp1}
        \begin{fmfgraph*}(140,60)\fmfkeep{comp1}
          \fmfpen{0.7thick}
          \fmfleft{i1,i2}\fmfright{o1,o2}
          \fmfv{l=$e$}{i1}\fmfv{l=$e$}{i2}\fmfv{l=$\ell$}{o2}\fmfv{l=$\ell$}{o1}
          \fmf{fermion,tension=1.1}{i1,v1}
          \fmf{fermion,tension=1.1}{v1,i2}
          \fmf{zigzag,label=$\varphi$,label.side=left,tension=0.9}{v1,v2}
          \fmf{fermion,tension=1.1}{v2,o1}
          \fmf{fermion,tension=1.1}{o2,v2}
          \fmfvn{decor.shape=circle,decor.filled=full,decor.size=1.7thick}{v}{2}
        \end{fmfgraph*}
      \end{fmffile}
    \end{tabular}
    \begin{tabular}{c}
      $\Rightarrow$
    \end{tabular}
    %\end{adjustbox}  & \adjustbox{raise={3ex}{\height}}{ $\Rightarrow$} &
    %\displaystyle\begin{adjustbox}{max size={40mm}{20mm},raise= {-1mm}{\height}}
    \begin{tabular}{c}
      \begin{fmffile}{compef}
        \begin{fmfgraph*}(140,60)\fmfkeep{compef}
          \fmfpen{0.7thick}
          \fmfleft{i1,i2}\fmfright{o1,o2}
          \fmfv{l=$e$}{i1}\fmfv{l=$e$}{i2}\fmfv{l=$\ell$}{o2}\fmfv{l=$\ell$}{o1}
          \fmf{fermion,tension=1.1}{i1,v1}
          \fmf{fermion,tension=1.1}{v1,i2}
          \fmf{fermion,tension=1.1}{v1,o1}
          \fmf{fermion,tension=1.1}{o2,v1}
          \fmfvn{decor.shape=circle,decor.filled=shaded,decor.size=7thick}{v}{1}
        \end{fmfgraph*}
      \end{fmffile}
      %\end{adjustbox}
      \end{tabular}
    \end{tabular}
  \end{figure*}

We now write the neutral mediator models in the form of 
\eqref{eq:compeq} for ease of comparison. When the mediator is a vector 
$V^{\mu}$ we have the interaction 
\begin{align}
  \mathcal{L}
  &=\frac{\lambda}{2}
  \bar{\ell}\gamma_{\mu}(\cos\phi+\sin\phi\gamma^5)\ell
  V^{\mu} 
\end{align}
where $\lambda$ is a real parameter and $\phi$ a mixing 
angle that parametrizes the relative strengths of vector and 
axial-vector interactions. This leads to the 
effective operator
\begin{align}
  \mathcal{O}_V
  &=-\frac{\lambda^2}{4m_V^2}
  \left[
  (1-\sin 2\phi)\bar{e}_L\gamma^{\mu} e_L
  \bar{\ell}_L\gamma_{\mu}\ell_L
  +
  (1+\sin 2\phi)\bar{e}_R \gamma^{\mu} e_R
  \bar{\ell}_R\gamma_{\mu}\ell_R
  \right.\nonumber\\ &\quad \left.
  +
  \cos 2\phi\,\bar{e}_L\gamma^{\mu} e_L
  \bar{\ell}_R\gamma_{\mu}\ell_R 
  +
  \cos 2\phi\,\bar{e}_R\gamma^{\mu} e_R
  \bar{\ell}_L\gamma_{\mu}\ell_L 
  \right].
\end{align}

We compare the above operator with the bounds given in 
\cite{Schael:2013ita} for the pure vector $\phi=0$ and pure axial 
vector $\phi=\tfrac{\pi}{2}$ cases. The 
combined bounds from all leptons (assuming flavor universality),
$e^+e^-\rightarrow\ell^+\ell^-$, leads to the most stringent 
constraints. The sign of the four-lepton operator
coefficient is relevant due to its interference with the SM
background.
The coefficient of all the operators that we consider
is negative, hence the $\Lambda^{-}$
bounds reported in \cite{Schael:2013ita} are relevant, and are
shown in table \ref{tab:compositeness}.

\begin{table}
  \centering
  \begin{tabular}{|c|cccc|c|}
    \hline
    \hline
    Bound &   \multicolumn{4}{c|}{Operators} & Limit  \\
    \cline{2-5} 
    & $\eta_{LL}$ & $\eta_{RR}$ & $\eta_{LR}$ & $\eta_{RL}$ & $\Lambda$ (TeV)\\
    \hline
    VV                & -1 & -1 &-1&-1& 20.0\\
    AA  & -1 & -1 & 1& 1& 18.1\\
    %RR & $\frac{\pi}{4}$ &  0 &  -1 & 0& 0&  11.3\\
    %LL & $\frac{3\pi}{4}$&  -1 &  0 & 0& 0& 11.8\\
    LR+RL &  0 &  0 & -1& -1& 14.5\\
    \hline
  \end{tabular}
  \caption{Summary of compositeness bounds on various four-lepton
  operators from LEP \cite{Schael:2013ita} arising from a neutral
  spin-1 and spin-0 mediators. Shown are the bounds for
  various combination of operators.
  In all cases the reported $\Lambda^-$ bounds apply.
  }
  \label{tab:compositeness}
\end{table}

For a spin-0 mediator $\varphi$, we treat the scalar and pseudoscalar 
interactions separately. The interaction Lagrangians are
\begin{align}
  \mathcal{L}_S
  &=
  \lambda
  \bar{\ell} \ell \varphi
  &
  \mathcal{L}_{PS}
  &=
  i\lambda\bar{\ell}\gamma^5\ell\varphi
  ,
\end{align}
where $\lambda$ is, in general, complex. The only resulting effective
operator which fits the form of \eqref{eq:compeq} is identical in
both cases,
\begin{align}
  \mathcal{O}_{S/PS}
  &=
  -\frac{2|\lambda|^2}{8m_{\varphi}^2}
  \left[
  \bar{e}_R\gamma^{\mu}e_R
  \bar{e}_L\gamma_{\mu}e_L
  \right].
\end{align}
Thus, for both the scalar and pseudoscalar
interactions we can use the combined Bhabha scattering bound from the
recent results \cite{Schael:2013ita} (see 
also \cite{Bourilkov08}).

In obtaining these limits we have assumed that the mass of the neutral 
mediator is larger than the center-of-mass energy of the collisions. 
Therefore, if the mediator is lighter than the LEP CM energy, 208 GeV, 
the bounds obtained from compositeness are not directly applicable.
\subsubsection{Resonant production at LEP}

When the mass of the mediator is below the LEP threshold, 208 GeV,
it can be resonantly produced. It can subsequently decay into
pair of leptons with a characteristic Breit-Wigner resonance shape in
the lepton pair invariant mass distribution. 
These bounds can be ameliorated by forcing the dominant decays of the
mediator to invisible or otherwise hidden final states. This typically
requires introducing significant complexity to the new physics model.

There are no available model independent studies for a leptophilic $Z'$ 
model at LEP. Limits exist in the context of specific models, which 
assume a specific pattern of couplings~\cite{Schael:2013ita}. A $Z'$ of 
mass less than 209 GeV is ruled out unless the coupling is less than 
about $10^{-2}$. A general analysis is likely to lead to a much stronger 
limit. However, this conservative limit itself will be seen to be the 
strongest constraint for mediator masses as low as a few GeV.

\subsubsection{Light neutral mediators}
If the neutral mediator is very light, then even a small coupling with
the leptons can give observable effects in $(g-2)_\mu$. Such light
mediators have a variety of constraints, which we briefly review here.
For a more detailed treatment of this topic, we refer to a recent review
\cite{Essig:2013lka}.

High intensity electron beams are dumped on to fixed targets. The SM
particles are stopped in a shield, but very weakly interacting neutral
mediators can escape this shield and their decays to pair of leptons
can be observed. The beam dump experiments at Fermilab and SLAC
\cite{Riordan:1987aw,Bjorken:1988as,Bross:1989mp} have been reanalyzed
in the context of light neutral
mediators \cite{Bjorken:2009mm}. The constraints apply for a range of
couplings -- small enough to escape the shielding but large enough to
yield an appreciable decay rate.

High luminosity $e^+ e^-$ colliders can produce the neutral mediator
radiatively $e^+ e^- \to \phi \gamma$, where $\phi$ is the neutral
mediator~\cite{Borodatchenkova:2005ct,Fayet:2007ua,Bjorken:2009mm}.
The decay $\phi
\to \mu^+ \mu^-$ provides the strongest constraint for mediator masses
between several hundred MeV and a few GeV.  We use the
reinterpreted constraints derived from BaBar data performed in
\cite{Reece:2009un}, which was carried out for a vector mediator. For
the same value of coupling, the spin-0 mediator production cross
section within the acceptance region is lower by roughly a factor of
two. We scale the limits on the coupling of the spin-0 neutral mediator 
appropriately to reflect this
fact.

The electron $g-2$ measurement is also sensitive to light neutral
mediators and provides the strongest bounds on this scenario for
mediator masses between a few MeV and several hundred MeV.
A number of proposed experiments aim to probe the region
of weakly coupling light mediators in the future.

\subsection{Charged mediators}
\subsubsection{Monophoton searches at LEP}
For DM masses less than half of the LEP center of mass (CM) energy, 
pairs of DM particles can be produced in association with a photon, $e^+ 
e^- \rightarrow \chi \chi \gamma$.

The existing monophoton analyses focus only on fermionic DM,
with the mediator heavier than the LEP threshold.
Thus, these analyses put constraints on 
four-fermion contact 
operators involving the electron and DM fields using
experimental limits on the process 
$e^+ e^- \rightarrow \chi \chi \gamma$.
The present limits in 
the literature only apply to a restricted set of operators. For the
simplified models that we will consider, the following 
Fierz rearrangements will be relevant,
\begin{align}
  \bar{\mu}(1+\gamma^5)\chi&\bar{\chi}(1-\gamma^5)\mu= \nonumber\\
  &\frac{1}{2}\left[
  \bar{\mu}\gamma^{\mu}\gamma^5\mu\bar{\chi}\gamma_{\mu}\gamma^5\chi-
  \bar{\mu}\gamma^{\mu}\mu\bar{\chi}\gamma_{\mu}\chi -
  \bar{\mu}\gamma^{\mu}\mu\bar{\chi}\gamma_{\mu}\gamma^5\chi
  +\bar{\mu}\gamma^{\mu}\gamma^5\mu\bar{\chi}\gamma_{\mu}\chi\right], \nonumber \\
  \bar{\mu}\gamma^{\mu}(1+\gamma^5)&\chi\bar{\chi}\gamma_{\mu}(1+\gamma^5)\mu=
  \nonumber\\
  &\bar{\mu}\gamma^{\mu}\gamma^5\mu\bar{\chi}\gamma_{\mu}\gamma^5\chi+
  \bar{\mu}\gamma^{\mu}\mu\bar{\chi}\gamma_{\mu}\chi +
  \bar{\mu}\gamma^{\mu}\mu\bar{\chi}\gamma_{\mu}\gamma^5\chi
  +\bar{\mu}\gamma^{\mu}\gamma^5\mu\bar{\chi}\gamma_{\mu}\chi.\label{eq:Fierz}
\end{align}

Of the four contact operators that appear after these rearrangements 
only the first two (vector-vector and axial vector-axial vector) are 
considered in the LEP analysis. We therefore use the more stringent of 
these two (the vector-vector term) for Dirac DM. In 
essence, we are assuming that interference between the terms will
not affect the bound significantly. For Majorana DM the
vector bilinear vanishes identically and so we use the axial 
vector-axial vector bound.

\subsubsection{Collider limits on charged mediators} 

Current collider limits exist only for scalar mediators. 
LEP rules out charged mediators lighter than about a 100 GeV.
The charged scalar mediators can also be pair-produced at the LHC, each of
which then decay to a lepton and DM. The signature is then a
pair of opposite-sign, like-flavor leptons with missing transverse
energy.  After Run 1, LHC puts constraints on the  production of
charged scalars. The relevant bounds are presented in the context of
supersymmetry, for pair production of sleptons -- charged scalars
carrying flavor. These analyses directly apply to charged scalar
mediator case \cite{ATLAS-CONF-2013-049}.  The relevant bounds for the
models we consider are the ``right-handed slepton'' bounds, where the
scalar does not carry SU(2) quantum numbers. In the case where the
scalar mediator carries SU(2) charge is more strongly constrained
due to its enhanced production cross section. The bounds also depend
on the DM mass, and get weaker in the region where the dark
matter mass is close to the mediator mass due to reduced missing
transverse energy.

\section{Analysis of Simplified Models \label{sec:models}}

In this section we consider specific simplified models of leptophilic
DM.  We classify the different theories based on the spin of the DM,
the spin and charge of the mediator, and the form of the interaction.
A similar approach for couplings of DM to quarks was adopted
in~\cite{DMclass2010,Fan:2010gt}, and more recently
in~\cite{Chang:2013oia,Bai:2013iqa,DiFranzo:2013vra}.

For each simplified model the parameter space can be separated into 
regions favored or disfavored by their effects on $\Delta a_{\mu}$. When 
applicable, limits from direct detection and collider experiments are 
also considered. It is convenient to present the results in terms of the 
parameter $\Lambda$, defined as $m_{\rm Med}/\lambda$, where $m_{\rm 
Med}$ is the mediator mass and $\lambda$ the mediator couplings to 
leptons.

When the contribution to $a_{\mu}$ is negative the tension between 
theory and experiment is increased. In this scenario, we require the new 
physics contribution to $a_{\mu}$ be no larger than twice the 
uncertainty in $\Delta a_{\mu}$. In other words, we require that the 
``wrong sign" contributions be indistinguishable from the combined 
experimental and theoretical uncertainties at two standard deviations, 
$|\hat{a}_{\mu}| \leq 2 \delta a_{\mu}$. As $\delta a_{\mu}$ is reduced 
these bounds become stronger.

When the sign of $\hat{a}_{\mu}$ is positive the DM contribution improves the 
agreement between theory and experiment. In these cases we identify the 
region of parameter space that eliminates the discrepancy, so that 
$a_{\mu}$ agrees with the experimentally measured value to within $2 
\delta a_{\mu}$. Specifically, we shade regions between the 
$\hat{a}_{\mu}$ values of $44.7\times 10^{-10}$ and $12.7\times 
10^{-10}$. In these cases, as $\delta a_{\mu}$ is reduced these bands
become more narrow about their central values.

The collider limits on charged mediators depend sensitively on their 
transformation properties under the electroweak SU(2$)_{\rm L} \times$ 
U(1$)_{\rm Y}$ gauge group. In general, the mediator and DM candidate 
could arise from any of several SU(2) representations, or even arise as 
an admixture of different representations.  In this section, when 
considering simplified models with charged mediators, we restrict our 
analysis to those theories where the mediator and DM carry no charge 
under SU(2$)_{\rm L}$. Under this assumption chiral symmetry violating 
couplings of the DM to muons are forbidden. The only source of violation 
of the chiral symmetry is then the muon mass.  Further, it allows us to 
limit our analysis to simplified models where the DM couples only to the 
right-handed muons. In the subsequent formulae we will set $a = 
\frac{\lambda}{2}$, $b = \pm \frac{\lambda}{2}$, with the choice of sign 
corresponding to the right-handed chiral projectors for muons.  Then, it 
can be seen from equations \eqref{eq:avf}-\eqref{eq:afs} that each of 
the charged mediator $\hat{a}_{\mu}$ contributions is a function of 
three dimensionless parameters, namely $\varepsilon,r$ and $\lambda$, 
and appears with a definite sign.

In the case of neutral mediators, the contribution to $a_{\mu}$ depends 
only on the coupling of the mediator to the muon, and is independent of 
the form of the interactions of the mediator with DM. The limits on 
neutral mediators from the LEP compositeness bounds are also independent 
of the coupling to DM. In general, we find that the compositeness bounds 
are much more restrictive than the monophoton bounds. Therefore, except 
in those cases for which the direct detection bounds are relevant, the 
limits on scenarios with a neutral mediator are independent of the 
details of the DM candidate. We now turn to analysing each simplified 
model in detail.

\subsection{Real Scalar Dark Matter\label{ssec:rsDM}}
\subsubsection{Charged Mediator\label{sssec:rsDMt}}
We begin with the case of a real scalar DM candidate $\chi$ and a charged mediator $F$, which is a fermion. The relevant interaction is
\begin{align}
  \mathcal{L}
  &=
  \frac{\lambda}{2}\bar{\mu}
  \left(1-\gamma^5 \right) F\chi +\;\text{h.c.}
\end{align}
Employing \eqref{eq:afs} to obtain the contribution to $a_{\mu}$ we find
\begin{align}
  \frac{1}{\Lambda^2}
  &\equiv
  \left(\frac{|\lambda|}{m_F}\right)^2
  =
  \frac{96\pi^2(1-r)^4\hat{a}_{\mu}^{F,S}}
  {m_{\mu}^2(1-6r^2+ 3r^4+2r^6-6r^4\ln r^2)}
\end{align}
with $r=m_{\chi}/m_F$.

\begin{figure}[tp]
\centering
\centering
\includegraphics[width=0.45\textwidth]{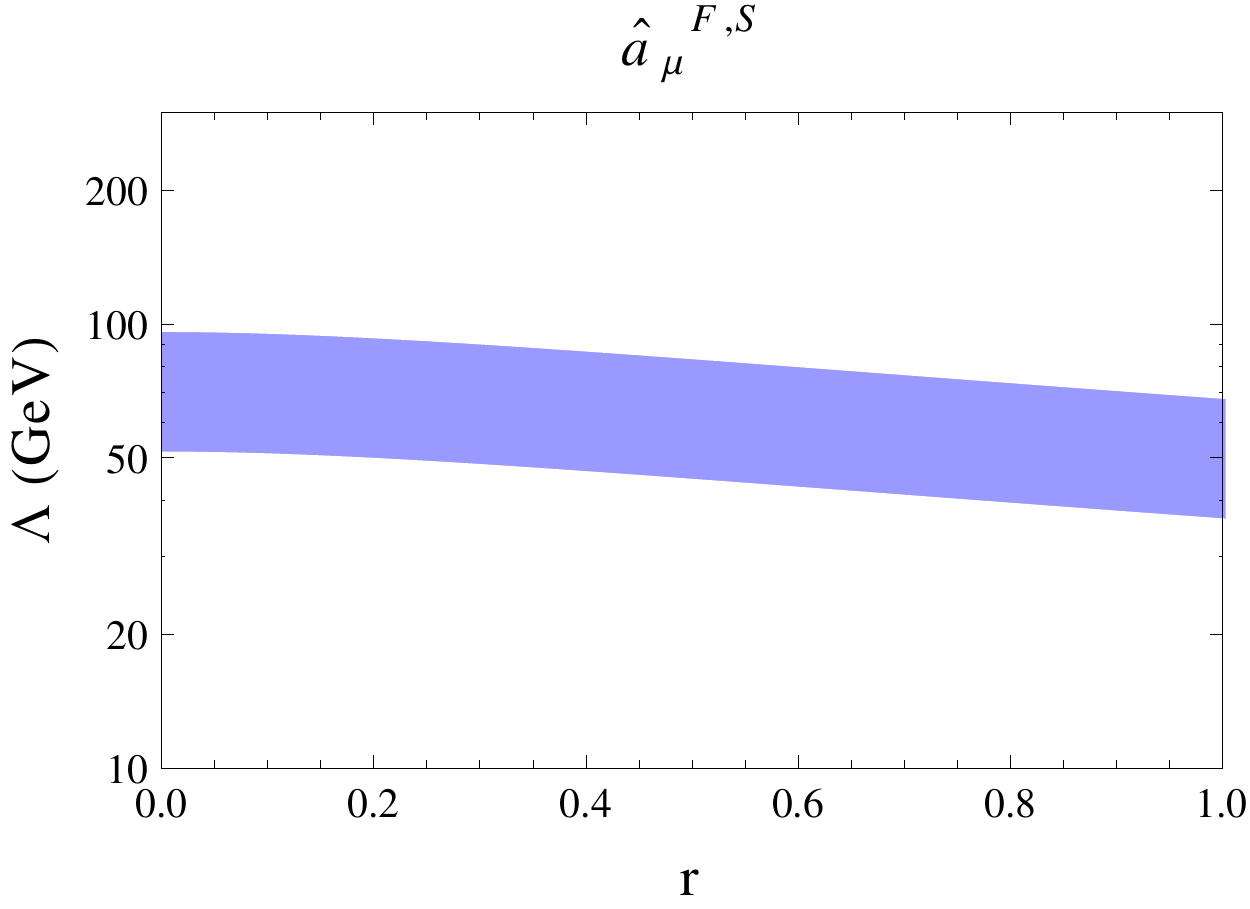}
\caption{We see the favored band (in blue) for real
scalar DM mediated by a charged fermion $F$ as a function of
$r=m_{\text{DM}}/m_F$. 
}
\label{fig:aFS1}
\end{figure}

In Fig.~\ref{fig:aFS1} we plot the region of parameter space for 
which this model removes the $g-2$ anomaly to within 2$\sigma$. This 
scenario is not significantly constrained by the current limits from 
direct detection. The monophoton analyses in the literature have only 
been performed for fermionic DM, and are not directly applicable to this 
model. 
From the plot we see that for small values of 
$r$, the preferred values of $\Lambda$ lie between 50 and 95 GeV. 
For $r$ values closer to 1 the preferred range runs from 
about 35 to 65 GeV.

The gauge quantum numbers of the fermion $F$, and the associated 
collider signals, are identical to those of a slepton in supersymmetric 
theories. The limits on the slepton mass from direct slepton pair 
production at LEP stand at about 100 GeV. There exist stronger bounds 
from the LHC for light DM masses~\cite{ATLAS-CONF-2013-049}. Given that 
the pair production cross section is generally larger for fermions than 
for bosons, it is quite likely that the limits on the mass of $F$ are at 
least at the same level. This would imply that the region of parameter 
space where the $g - 2$ anomaly is explained is disfavored by LEP, 
unless the coupling $\lambda$ is large, $\lambda \gtrsim 1$. However, a 
more careful analysis is required in order to validate this conclusion. 
We leave this for future work.

\subsubsection{Neutral Spin-0 Mediator\label{sssec:rsDMss}}

We now turn to the case of a real spin-0 neutral mediator, $S$.  A few 
comments are in order. Clearly, a coupling of the form $\bar{\ell} \ell 
S$ is not consistent with gauge invariance under SM SU(2) if $S$ is a 
singlet. In principle, this problem can be avoided if $S$ arises as a 
linear combination of the SM Higgs and a singlet, but the theory would 
no longer be leptophilic since $S$ would also couple to quarks.

A potential UV completion is to add to the theory a heavy scalar SU(2) 
doublet $\widehat{H}$ which has no VEV, and couples only to leptons. After 
electroweak symmetry breaking the neutral components of $H$ can mix with 
a real singlet scalar $\widehat{S}$ via a coupling with the SM Higgs $H$. A 
light state $S$ with couplings to leptons can then emerge as a linear 
combination of the singlet $\widehat{S}$ and the neutral components of the 
doublet $\widehat{H}$. The relevant interactions have the schematic form
 \begin{equation}
(\hat{\lambda} \bar{L} \widehat{H} E + \kappa \widehat{S}
H^{\dagger} \widehat{H} +
\;\text{h.c.} ) + M^2 \widehat{H}^{\dagger} \widehat{H}
 \end{equation}  
 where $L$ is the SU(2) doublet SM lepton and $E$ the SM singlet. The 
$CP$ transformation properties of $S$ depend on whether the coupling 
$\kappa$ is real or complex.
We assume for concreteness that the 
coupling matrix $\hat{\lambda}$ is universal and flavor diagonal in the 
lepton mass basis.

The form of the low energy Lagrangian depends on the $CP$ properties of 
the scalar $S$. For real scalar DM, there is no
$CP$-conserving coupling of the DM with a single pseudoscalar at the
renormalizable level. Thus, we only consider the scalar mediator in
this case. The Lagrangian is of the form
 \begin{align}
  \mathcal{L}_S
  &=
  \lambda \bar{\mu}\mu S
\end{align}

Using \eqref{eq:as} we find the leading contributions to $a_{\mu}$ for
the scalar mediator. The contribution is written as
\begin{align}
    \left(\frac{|\lambda|}{m_S^2}\right)^2
    &=
    \frac{32\pi^2 \hat{a}^{S}}
    {m_{\mu}^2 
    \ln\left(\frac{m_S^2}{m_{\mu}^2} \right)}, 
\end{align}

\begin{figure}[tp]
\centering
\includegraphics[width=0.45\textwidth]{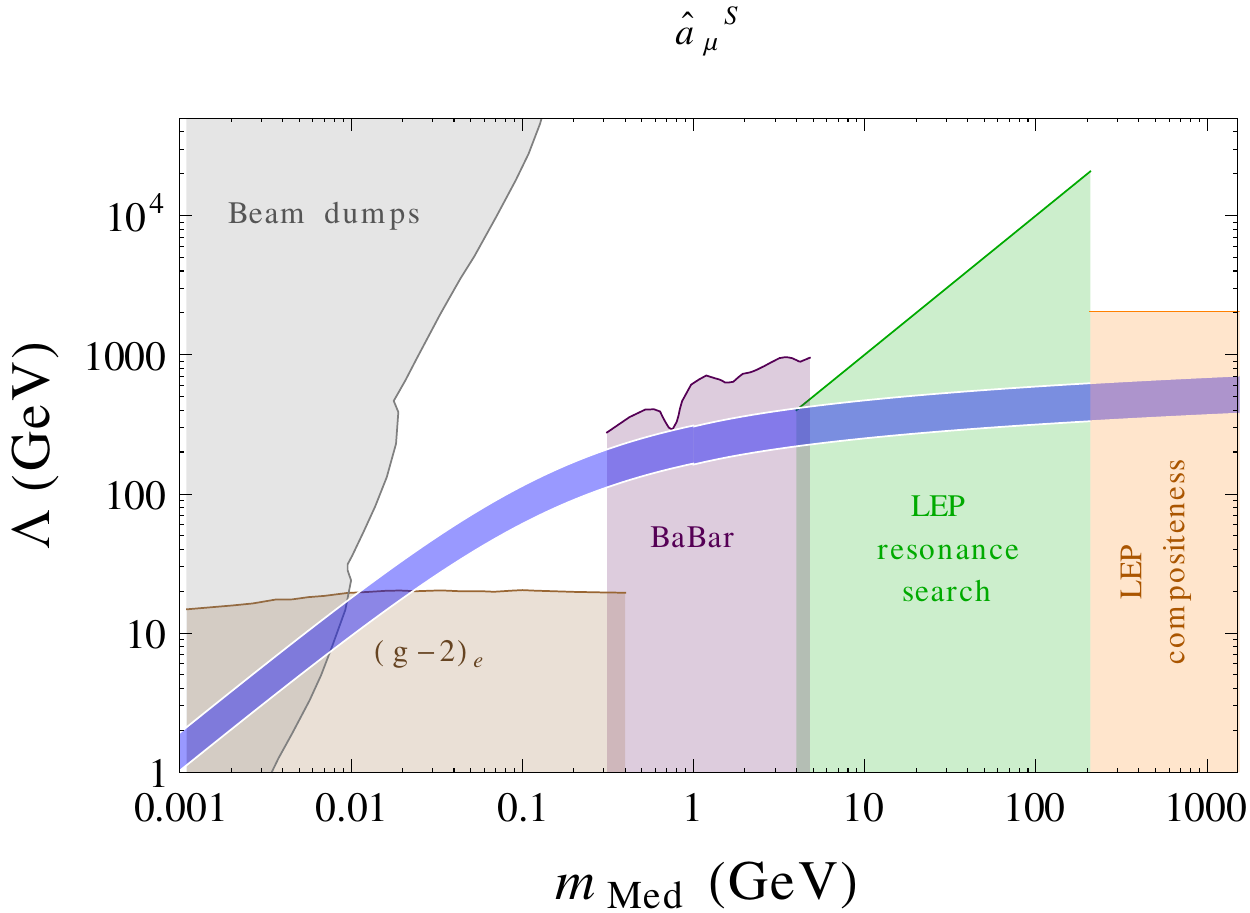}
\caption{\label{fig:aS1} Plots of favored regions of
parameter space for real scalar DM  with interactions
mediated by a neutral scalar $S$ as a function of $m_{Med}$.
The shaded blue
region corresponds to the range of parameters favored to set
$\Delta a_{\mu}=0$ to within 2$\sigma$. 
We also show bounds on light neutral scalars as colored shaded regions
(compositeness bounds, resonant production at LEP, BaBar and beam
dumps, and the electron $g-2$).}
\end{figure}

In Fig.~\ref{fig:aS1} we plot, over a range of mediator masses, the
region of
parameter space favored by the scalar interaction. We
also plot the corresponding compositeness bound from LEP for heavy
mediators. We also show bounds from resonant production at LEP and at
BaBar for lower mediator masses. We see from the figure that these
constraints rule out the region of parameter space where the scalar
interaction can account for the observed discrepancy in $a_{\mu}$,
except a small region of parameter space where the mediator is light,
between about 10 MeV and 300 MeV. This scenario is not significantly
constrained by current direct detection experiments.

The case of a neutral vector mediator need not be considered. The reason 
is that in a renormalizable theory, real scalar DM cannot couple to a 
spin-1 neutral mediator without violating $CP$. There is therefore no 
simplified model of this type that meets our criteria.

\subsection{Complex Scalar DM\label{ssec:csDM}}
\subsubsection{Charged Mediator\label{sssec:csDMt}}
 We proceed to complex scalar DM $\chi$. As before, we begin with a 
charged fermion, denoted by $F$, as the mediator. The interaction 
Lagrangian takes the form
 \begin{align}
\mathcal{L}
&=
\frac{\lambda}{2}\bar{\mu}\left(1-\gamma^5 \right) F\chi +\;\text{h.c.}
\end{align}
 As in the case of the real scalar \eqref{eq:afs} leads to
 \begin{align}
\left(\frac{|\lambda|}{m_F}\right)^2=\frac{96\pi^2(1-r)^4\hat{a}_{\mu}^{F,S}}{m_{\mu}^2(1-6r^2+ 3r^4+2r^6-6r^4\ln r^2)}
 \end{align}
where $r=m_{\chi}/m_F$.

The monophoton analyses in the literature do not apply to this model. 
However, as shown in section \ref{sec:compbnds}, this theory is 
constrained by direct detection. The parameter $\widetilde{\lambda}$ can 
be obtained by calculating the contributing diagram in the effective 
theory with the heavy mediator integrated out, in the leading logarithm 
approximation. Expressing the result in the form of \eqref{eq:diffxsec} 
this defines $\widetilde{\lambda}$ which is found to be
 \begin{align} 
\widetilde{\lambda}=\frac{\lambda^2\alpha}{12\pi^2m_{\phi}^2}\left[\ln 
\frac{m_{\tau}^2}{m_{\phi}^2}+ \ln \frac{m_{\mu}^2}{m_{\phi}^2}+ \ln 
\frac{|k|^2}{m_{\phi}^2} \right] 
 \end{align} 
 where $\alpha$ is the fine structure constant. In this calculation we 
have worked to zeroth order in the small parameter $k^2/m^2_{\ell}$, 
except in the case of the electron loop. For this diagram, we instead 
work to zeroth order in $m^2_{e}/k^2$, and this leads to the dependence 
on $|k|$ of the final result. When bounds are set we use the value 
$|k|=$30 MeV.

\begin{figure}[tp]
\centering
\includegraphics[width=0.45\textwidth]{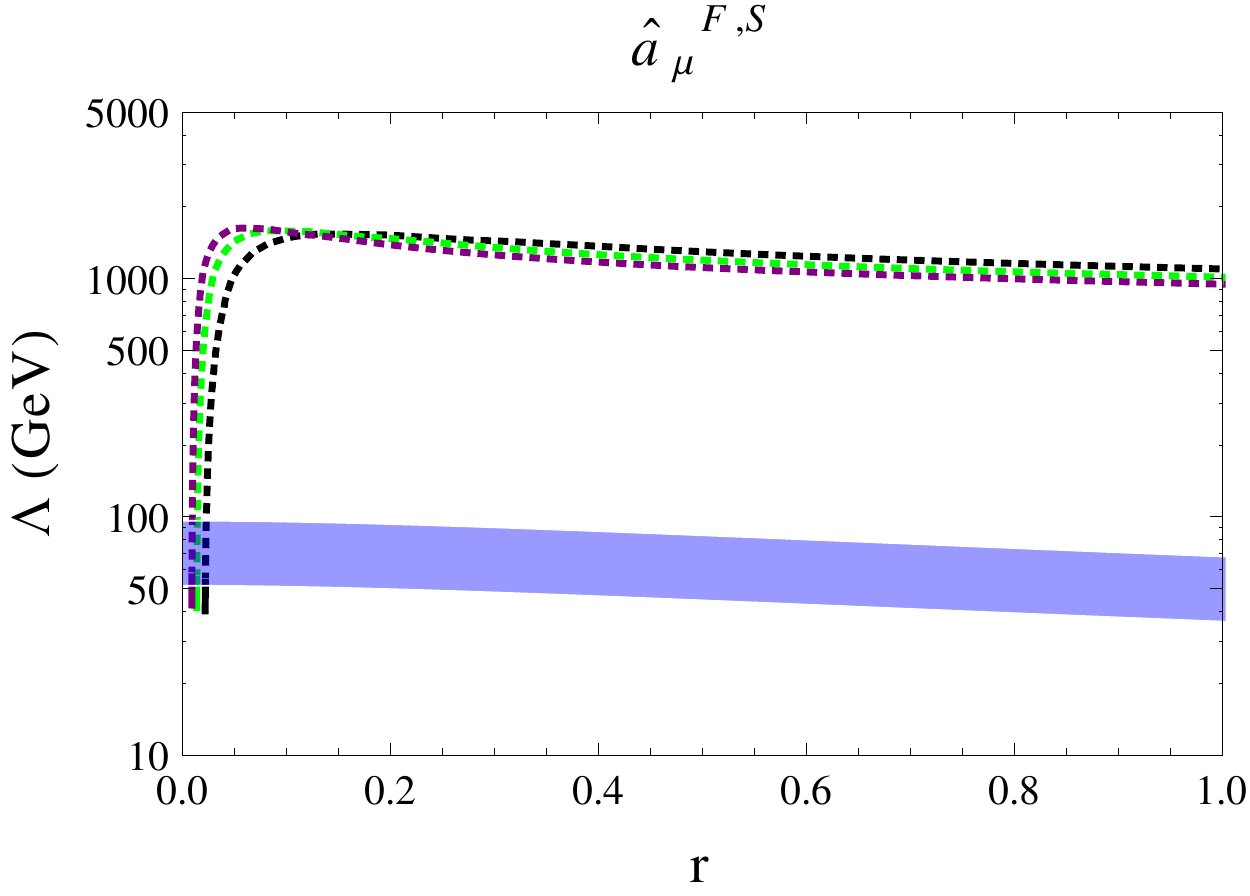}
\caption{\label{fig:csFc}Regions of favored parameter space compared
with bounds from direct detection and compositeness 
for a complex scalar DM mediated by a charged fermion.
The constraints are shown in terms of
$r=m_{\text{DM}}/m_{\text{Med}}$.
The direct detection bounds are shown as dotted lines, denoting
a 250 (black), 400
(green), or 600 GeV (brown) mediator mass. }
\label{fig:aFSdd1}
\end{figure}

In Fig. \ref{fig:aFSdd1} we have overlaid these bounds on the region of 
parameter space which for which this model removes the $g-2$ anomaly to 
within 2$\sigma$. The three bands shown in the figure correspond to 
mediator masses of 250 (black), 400 (green), and 600 GeV (brown). We see 
that the direct detection limits exclude almost the entire region, 
unless the DM mass lies below about 5 GeV. We further expect that a 
recast analysis will show that the limits on direct slepton searches at 
LEP disfavor the entire band, except for fairly large values of the 
coupling $\lambda$.

\subsubsection{Neutral Spin-0 Mediator\label{sssec:csclss}}
Next, we consider a spin-0 neutral mediator $S$. The interaction 
Lagrangian is of the form
\begin{align}
  \mathcal{L}_S
  &=
  \lambda\bar{\mu}\mu S
\end{align}
for scalar mediators. 

Since the direct detection constraints on this scenario are not 
significant, the analysis for neutral spin-0 mediators performed for the 
real scalar DM applies to this case as well. The results are shown in 
Fig.~\ref{fig:aS1}.

\subsubsection{Neutral Spin-1 Mediator\label{sssec:csclsv}}
In the case of a neutral spin-1 mediator $V^{\nu}$, the interaction 
Lagrangian takes the form
 \begin{align}
\mathcal{L}
&= \frac{i\lambda}{2}
\left(\chi^{\ast}\partial_{\nu}\chi-\chi\partial_{\nu}\chi^{\ast}
\right)V^{\nu} +\frac{\lambda}{2}\bar{\mu}\gamma^{\mu}\left(\cos\phi +\sin\phi\gamma^5 \right)\mu V_{\mu}
 \end{align}

Then \eqref{eq:av} leads to
\begin{align}
  \left(\frac{\lambda}{m_V}\right)^2
  &=\frac{16\pi^2 \hat{a}^V_{\mu}}
  { \displaystyle m_{\mu}^2\left(\cos 2\phi-\frac{2}{3} \right)} 
  \; . 
  \label{eq:schancont} 
\end{align} 
 Over the domain of $\phi$, this expression takes on both positive and 
negative values. We choose two specific cases, $\phi=0$, vector coupling 
and $\phi=\pi/2$, the axial vector coupling, as benchmarks. There are 
also additional constraints from direct detection. Axial vector 
mediators give no contribution, so these bounds only apply to the vector 
($\phi=0$) case. The DM-nucleon scattering amplitude can be expressed 
in the form of Eq. \eqref{eq:diffxsec}. In the leading logarithm 
approximation we obtain
 \begin{align}
  \widetilde{\lambda}
  &=
  \frac{\lambda^2 \alpha}{6\pi m_V^2} 
  \left[\ln\frac{m_{\tau}^2}{\Lambda_V^2} 
  +\ln\frac{m_{\mu}^2}{\Lambda_V^2} +\ln\frac{|k|^2}{\Lambda_V^2} \right].
\end{align}
 Here $\Lambda_V$ is the scale at which the logarithmic divergence is cut 
off. For concreteness, we take $\Lambda_V = 1$ TeV.

\begin{figure}[tp]
\centering 
\includegraphics[width=0.45\textwidth]{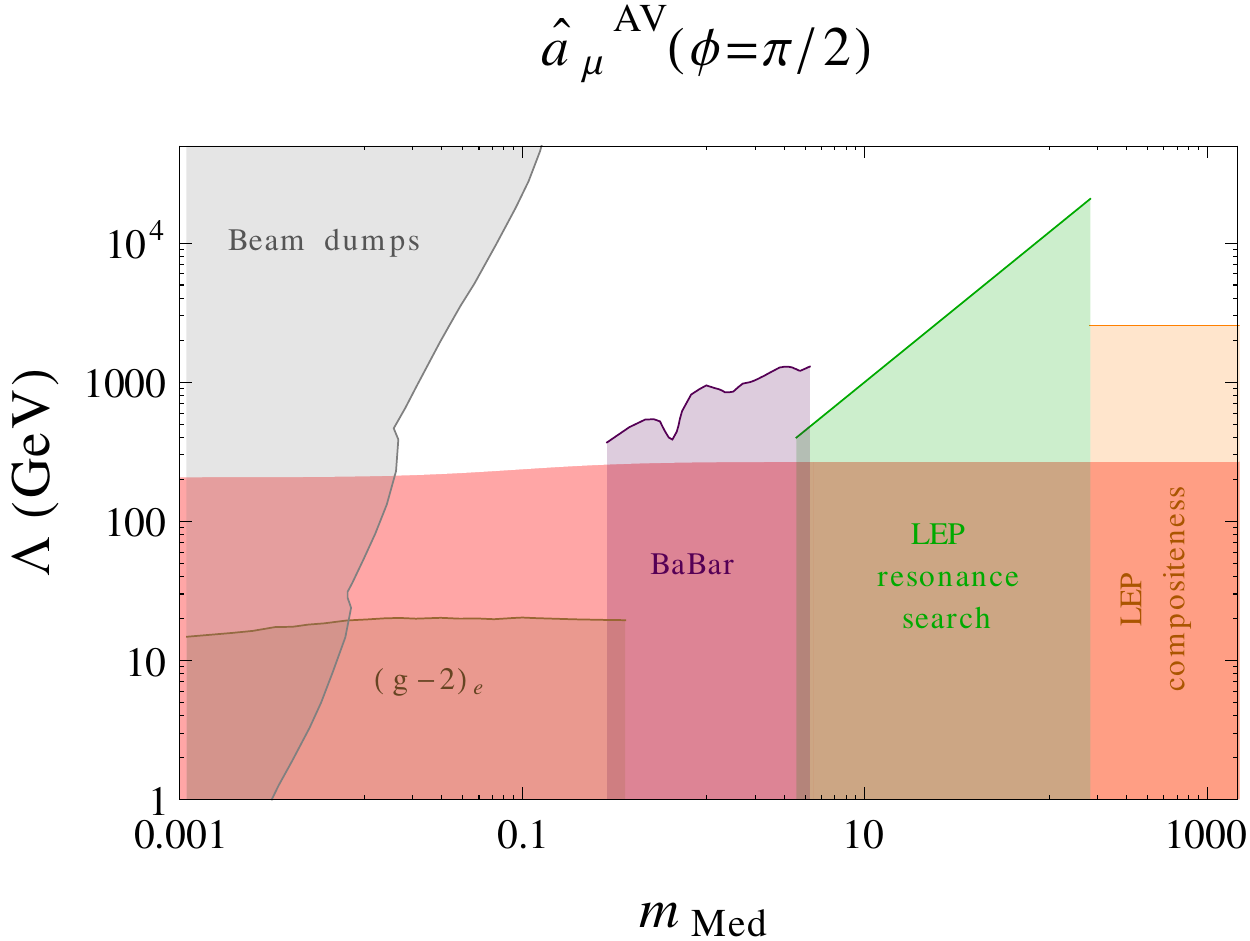}
\qquad 
\includegraphics[width=0.45\textwidth]{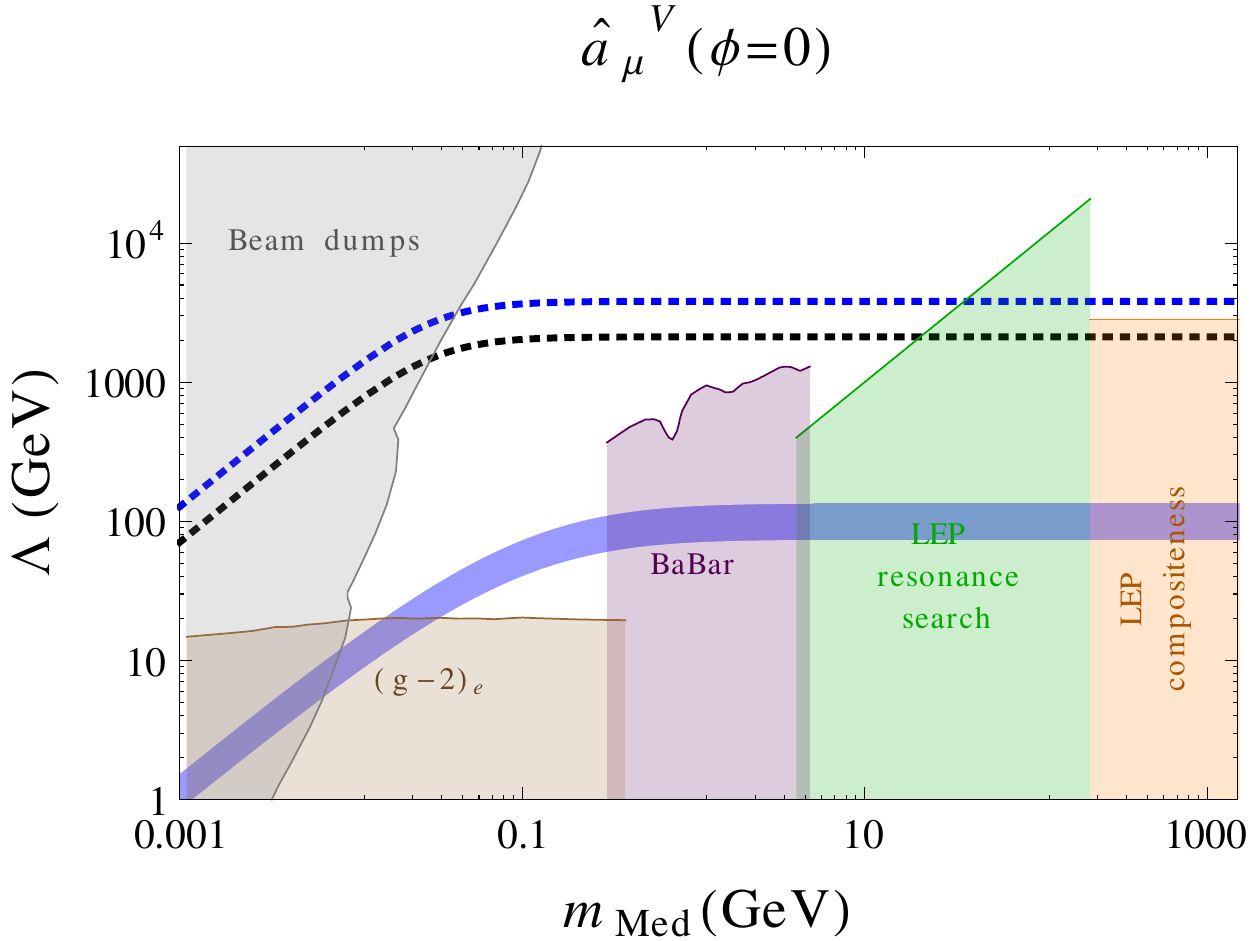}
\caption{ Limits on complex scalar DM mediated by a neutral spin-1
particle with vector interactions as a function of $m_{Med}$ for a
axial-vector mediator (left) and for a vector mediator (right).  For
the vector mediator, we also show direct detection bounds for two
different DM masses ($m_{DM}$=10 GeV (blue dotted) and 90 GeV
(black dotted)).}
\label{fig:aVdd1}
\end{figure}

In Fig.~\ref{fig:aVdd1} we show the favored band of parameter space 
for the vector case, and compare it to bounds from direct detection,
and from resonant collider production. 
We see that the entire region of parameter space that 
explains the anomaly in $a_{\mu}$ is excluded by direct detection
experiments if the DM particle is accessible to them
($m_{DM}\gtrsim 5$ GeV). Otherwise, a small window of mediator masses
between about 10 and 300 MeV is still allowed. 
For the axial-vector, the direct detection bounds do not apply,
therefore the muon $g-2$ measurement is the most sensitive in the
same mediator mass range.

\subsection{Majorana DM\label{ssec:mDM}}
\subsubsection{Charged Spin-0 Mediator\label{sssec:mDMt}}
Turning to fermionic DM we begin by considering a Majorana fermion 
$\chi$ whose interactions with leptons are mediated by charged scalars 
$\phi_l$. The interaction Lagrangian takes the form
 \begin{align}
   \mathcal{L}
   &=
   \frac{\lambda}{2}\bar{\mu}\left(1 -\gamma^5 \right)\chi \phi_l+\;\text{h.c.}
 \end{align}

\begin{figure}[tp]
\centering
\subfloat[ ]{
\centering
\includegraphics[width=0.45\textwidth]{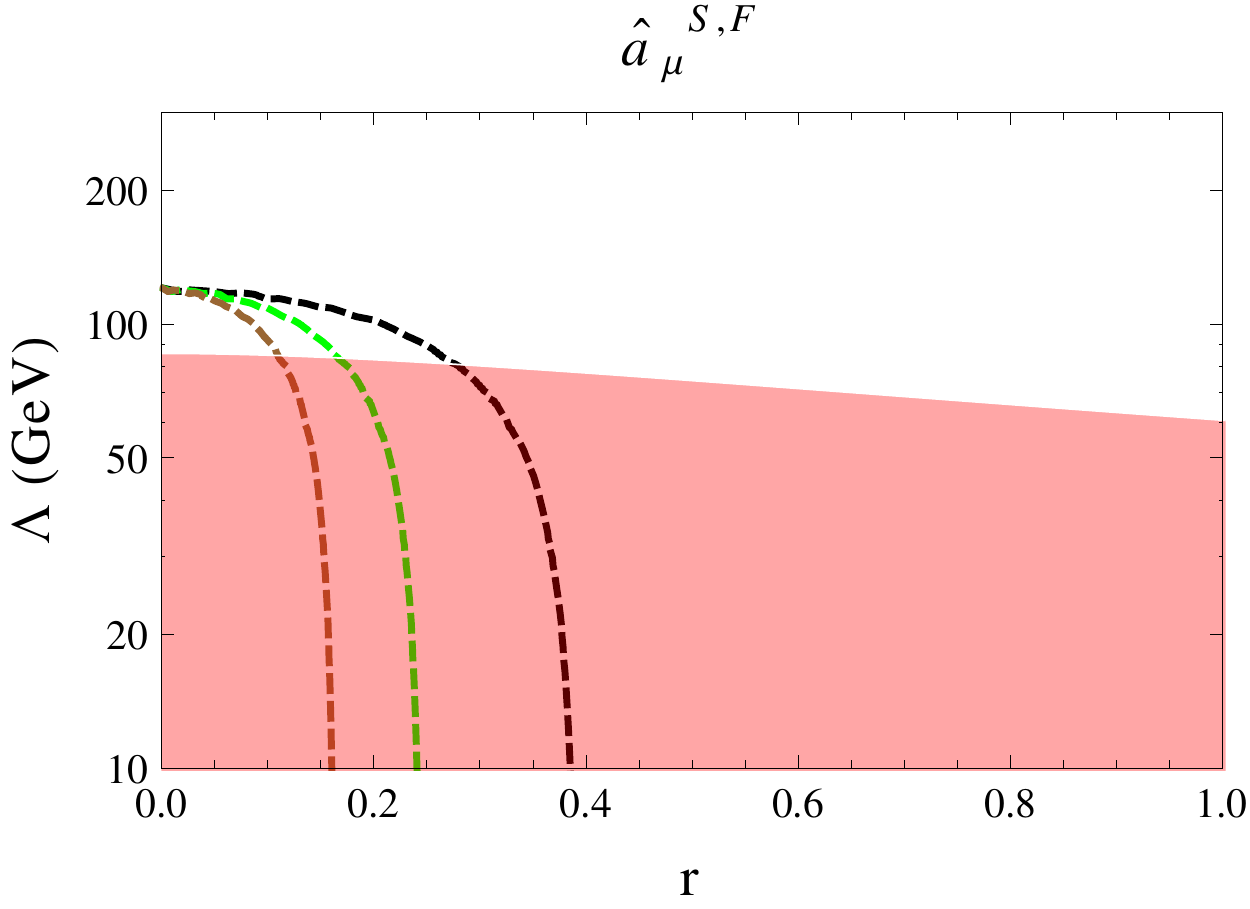}
\label{fig:asflepmf}
}
\qquad
\subfloat[ ]{
\centering
\includegraphics[width=0.45\textwidth]{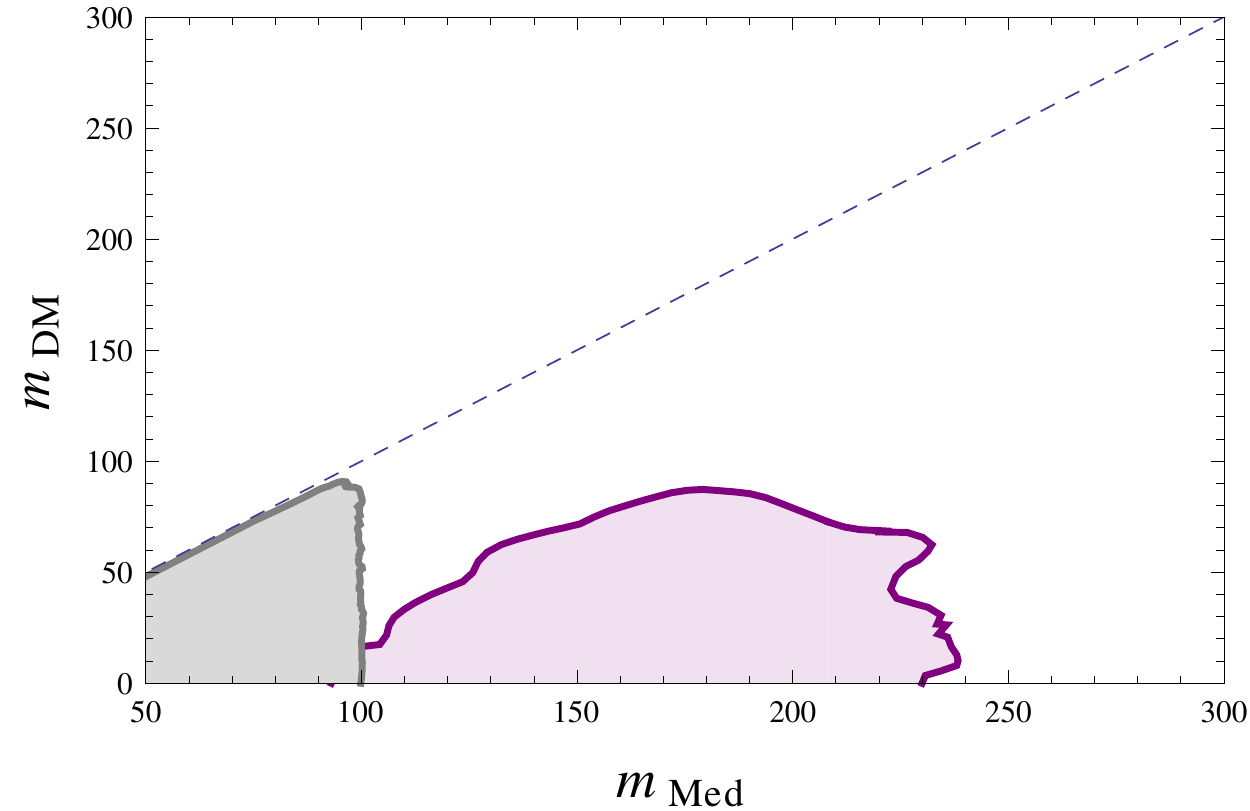}
\label{fig:slepsearch}
}
\caption{
(a)
The $g-2$ bound on Majorana DM with a charged
scalar mediator 
as a function of $r=m_{\text{DM}}/m_{\text{Med}}$. 
Dashed lines indicate the monophoton bounds from 
LEP, and are plotted for 250 (black), 400 (green), and 600 GeV
(brown) mediators.
(b) LHC (purple) and LEP (gray) constraints on charged scalars as a function of the
mediator and DM masses.
}
\end{figure}

The contribution to $a_{\mu}$ can be obtained from \eqref{eq:asf} and 
leads to the relation
\begin{align}
  \left(\frac{|\lambda|}{m_\phi^2} \right)^2
  =-\frac{96\pi^2 \hat{a}_{\mu}^{S,F}(1-r)^4} 
  {m_{\mu}^2(1-6r^2+3r^4+2r^6-6r^4\ln r^2)}
\end{align} 
 where $r=m_{\chi}/m_\phi$. The sign of the contribution increases the 
tension between theory and experiment. This implies that $a_{\mu}$ can be 
used to place limits on this model. The bounds on this model from direct 
detection are not expected to be significant. There are constraints, 
however, from the LEP monophoton searches. We compare these limits to 
those arising from $a_{\mu}$ in Fig. \ref{fig:asflepmf}. The three dashed 
lines correspond to 250 GeV (black), 400 GeV (green), and 600 GeV (brown) 
mediators.

From the figure we see that as the mass of the mediator increases 
the monophoton limits become weaker, and eventually become less 
sensitive than the bounds from $a_{\mu}$. The
$g-2$ bound restricts $\Lambda$ to be greater than 75 GeV when 
$r=0.5$ and greater than 85 GeV when $r=0.1$. Note, however, that the 
spin and gauge quantum numbers of $S$, and its decay modes, are 
identical to those of a slepton in supersymmetric theories. 
Therefore, the right-handed slepton bounds from LEP and the LHC apply,
and are shown in Fig. \ref{fig:slepsearch}.
For $\lambda=1$,
the lower bound on the slepton mass from LEP
and the LHC provides a stronger constraint on this model than 
$a_{\mu}$.

\begin{figure}[th]
\centering
\includegraphics[width=0.45\textwidth]{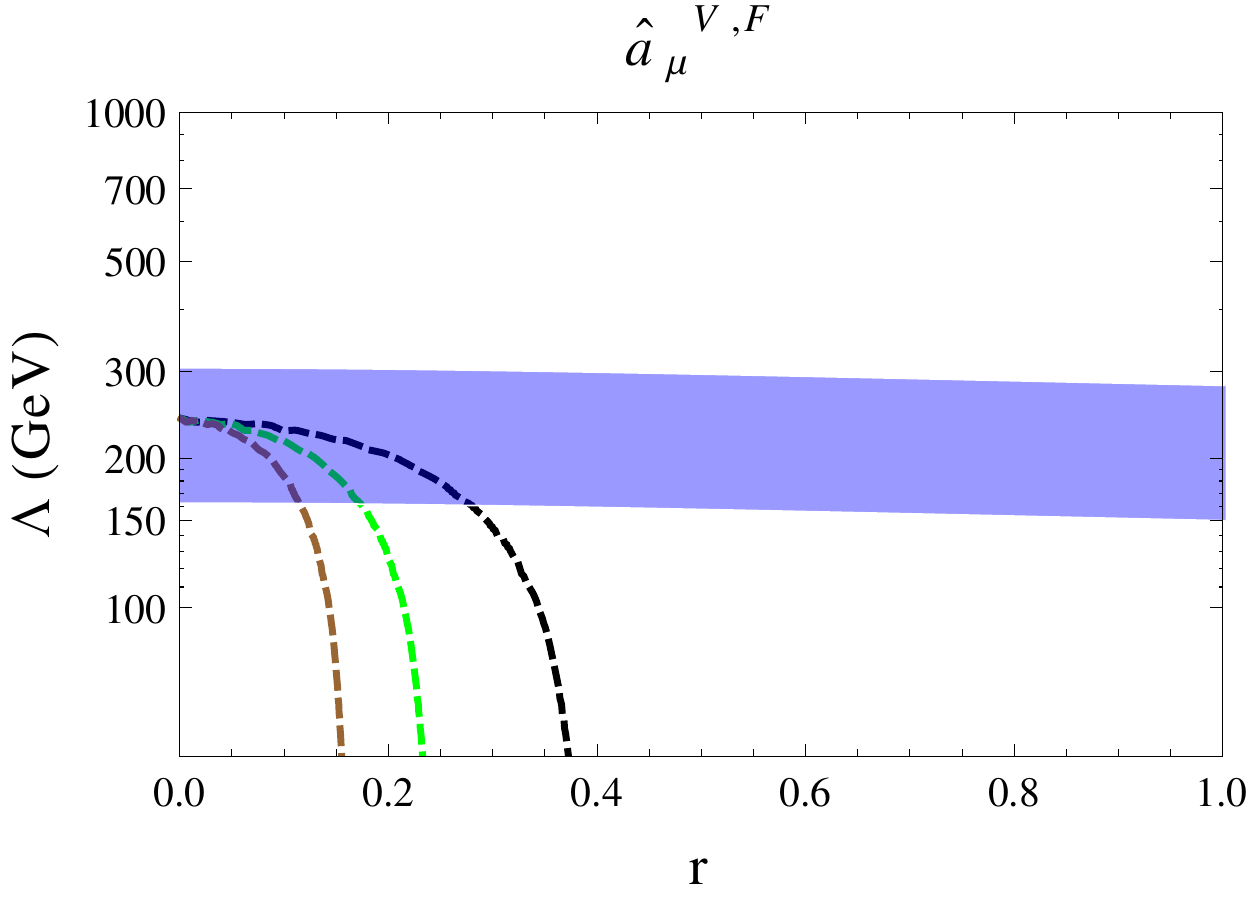}
\caption{
Constraints on Majorana DM with a charged spin-1 mediator
as a function of $r=m_{\text{DM}}/m_{\text{Med}}$. 
We show region favored to set $\Delta a_{\mu}=0$ to
within 2$\sigma$.
Dashed lines indicate the monophoton bounds from 
LEP, and are plotted for 250 (black), 400 (green), and 600 GeV
(brown) mediators.
}
\label{fig:aVFlepmf}
\end{figure}

\subsubsection{Charged Spin-1 Mediator\label{sssec:mDMtv}}
Next, we consider mediation by  charged vectors $V_l^{\nu}$. The interaction
Lagrangian is given by
\begin{align}
  \mathcal{L}
  &=
  \frac{\lambda}{2}\bar{\mu}\gamma_{\nu}\left(1+\gamma^5 \right)\chi
  V_l^{\nu}+\;\text{h.c.}
\end{align}

 The contribution to $a_{\mu}$ can be obtained from \eqref{eq:avf}, 
and leads to the relation
 \begin{align} 
\left(\frac{|\lambda|}{m_V^2} \right)^2=\frac{96\pi^2 \hat{a}_{\mu}^{V,F}(1-r)^4} {m_{\mu}^2(10-43r^2+78r^4-49r^6+4r^8+18r^6\ln r^2)} 
 \end{align} 
 where $r = m_{\chi}/m_V$. This expression can be used to determine
 the region of parameter space favored to remove the $g-2$ anomaly.
 While the direct detection bounds on this model are not expected 
 to be significant,
 there are constraints from LEP monophoton searches that we compare to 
 this
 favored region in Fig.~\ref{fig:aVFlepmf}. In the figure the LEP
 monophoton bounds are plotted for 250 GeV (black), 400 GeV (green),
 and 600 GeV (brown) mediators. We see from the plot that the values
 of $\Lambda$ between 165 and 285 GeV fall within the favored band.
 Within this range LEP bounds disfavor DM masses lighter than about 
 15 GeV, but there remains a significant region of parameter space that 
 can explain the discrepancy in $a_{\mu}$.

\subsubsection{Neutral Spin-0 Mediator\label{sssec:mDMss}}
When mediation occurs through a neutral spin-0 particle $S$, the 
Lagrangian takes the form 
\begin{align}
  \mathcal{L}_S
  &=
  \lambda
  \bar{\mu}\mu S
\end{align}
 We only consider the scalar mediator since there is no $CP$-conserving 
coupling of a pseudoscalar to a Majorana fermion at the renormalizable 
level. While the current limits from direct detection on leptophilic 
Majorana fermion DM are weak, the other limits on a neutral scalar 
mediator continue to apply. The results in Fig.~\ref{fig:aS1} again show 
the most stringent bounds.

\subsubsection{Neutral Spin-1 Mediator\label{sssec:mDMsv}}
\begin{figure}[tp]
\centering
\includegraphics[width=0.45\textwidth]{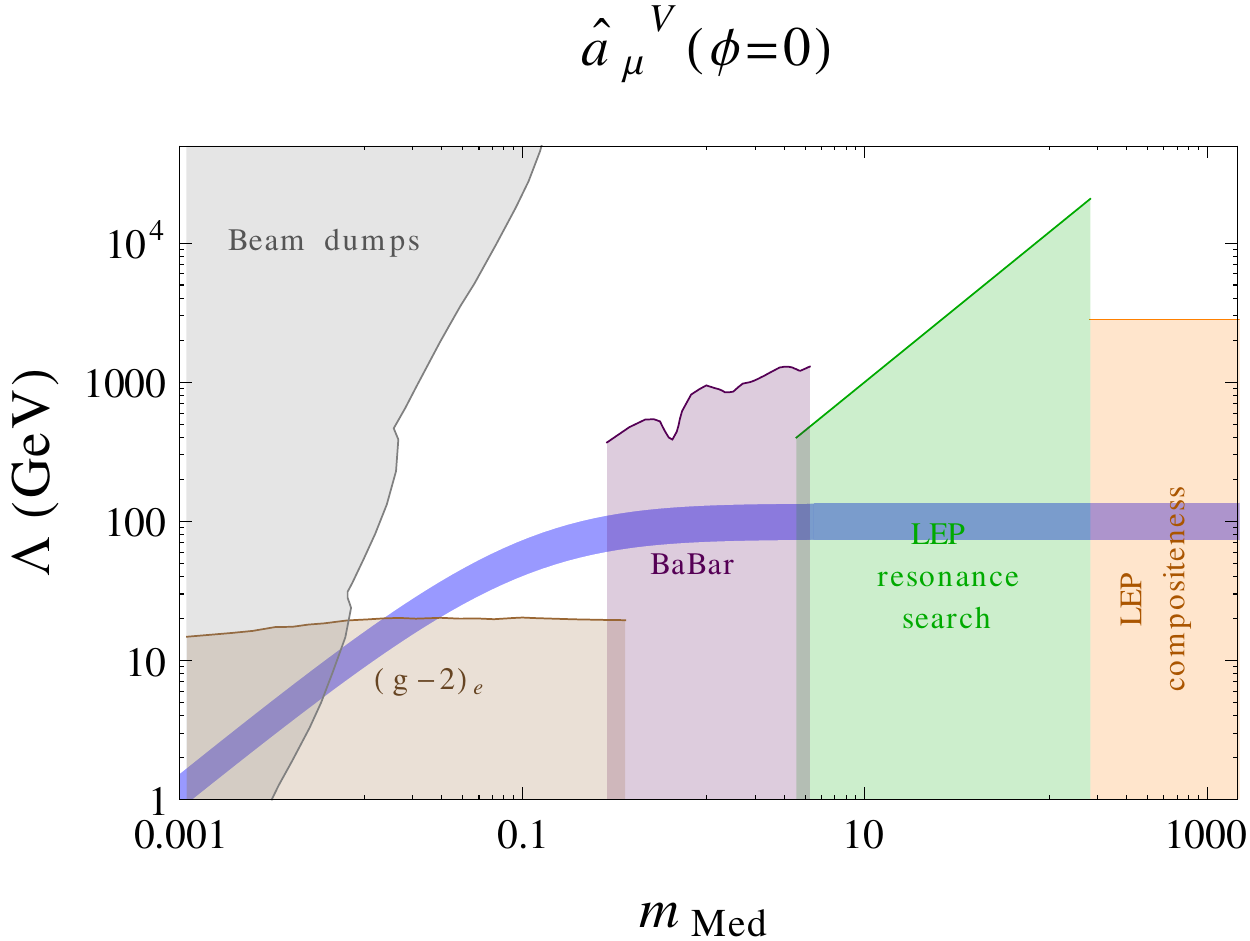}
\caption{ 
Shown are the 
favored values which set $\Delta
a_{\mu}=0$ to within 2$\sigma$ (blue band) for the neutral vector
mediator for Majorana DM. }
\label{fig:mdmV}
\end{figure}

 If the interaction is mediated by a neutral massive spin-1 particle 
$V^{\nu}$, we have for the interaction Lagrangian
\begin{align}
  \mathcal{L}&=
  \lambda_V\bar{\chi}\gamma_{\nu}\gamma^5\chi V^{\nu} +
  \frac{\lambda}{2}\bar{\mu}\gamma^{\mu}\left(\cos\phi +\sin\phi\gamma^5 \right)\mu V_{\mu}.
\end{align}
 The results are identical to those for a neutral spin-1 mediator 
coupling to complex scalar DM, except that the direct detection 
constraints from that case no longer apply. We plot the results in
Fig. \ref{fig:mdmV}. 
For axial-vector couplings, the results derived for complex scalar DM 
apply directly, and are shown in Fig. \ref{fig:aVdd1}(a).

\begin{figure}[tp]
  \subfloat[ ]{
\centering
\includegraphics[width=0.45\textwidth]{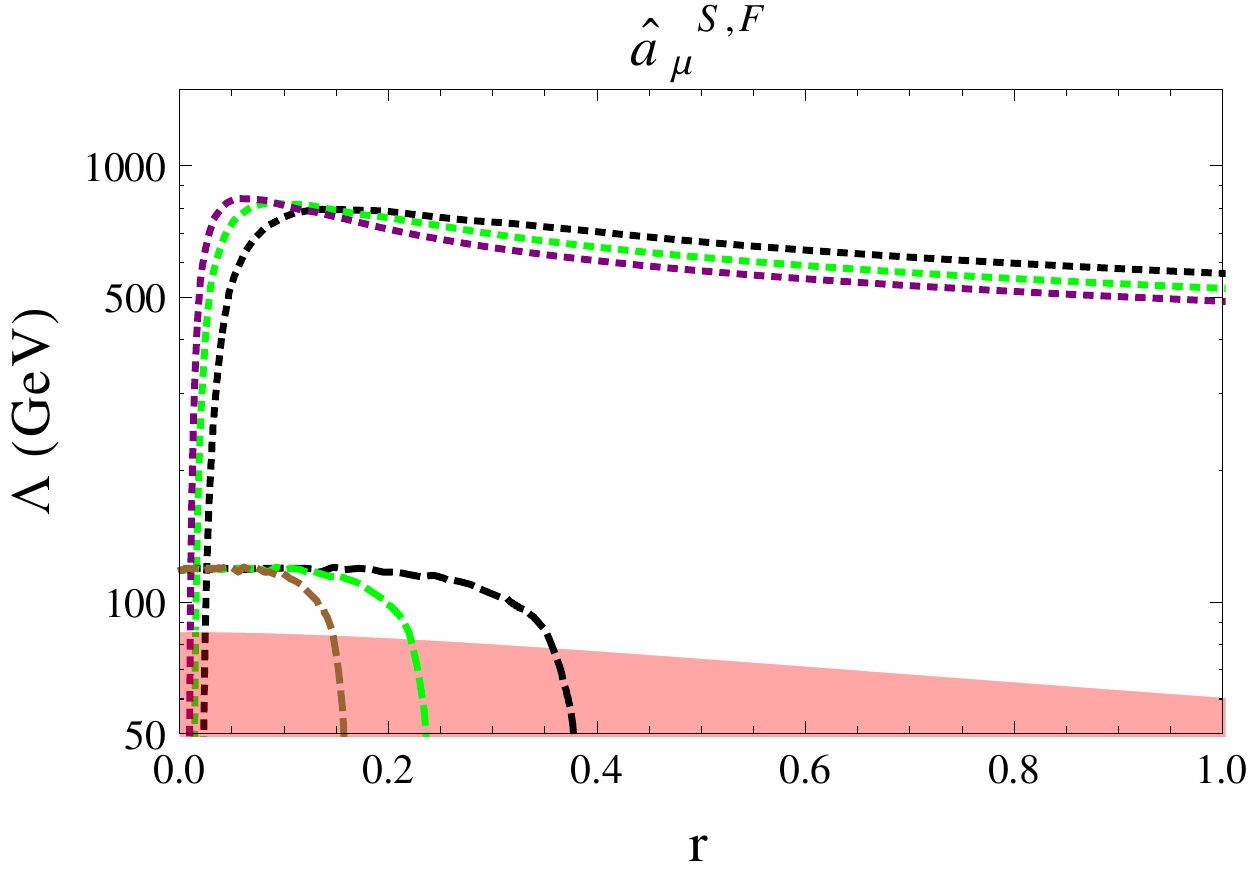}
\label{fig:aSF150all}
}
\qquad
\centering
\subfloat[ ]{
\centering
\includegraphics[width=0.45\textwidth]{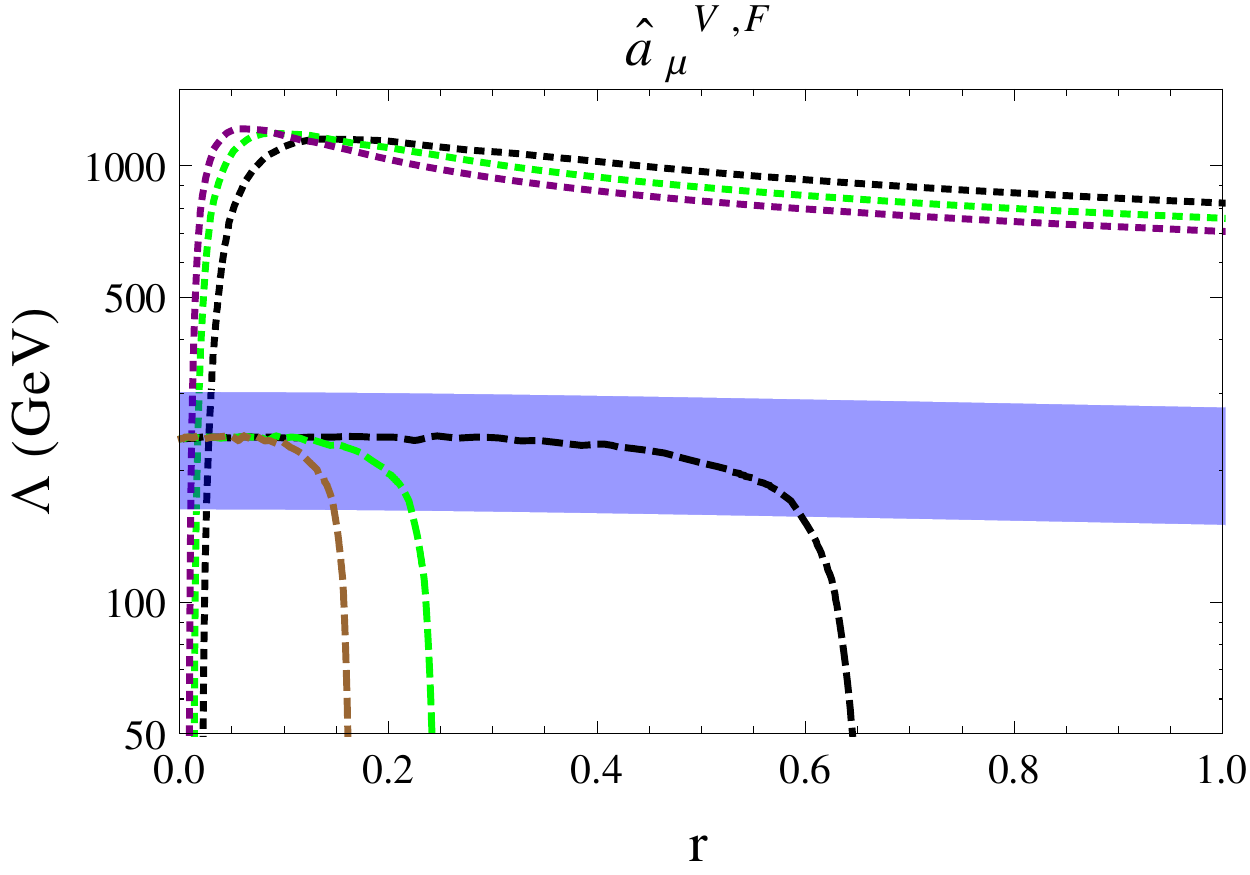}
\label{fig:aVF150all}
} 
\caption{\label{fig:mfAVdfSF} Plots as a function of
$r=m_{\text{DM}}/m_{\text{Med}}$ the $g-2$ analysis of Dirac DM with a
charged (a) scalar and (b) vector mediator. In (a) we plot the bound requiring
the contribution to $a_{\mu}$ be within 2$\sigma$ of 0. In (b) we show
the region of parameter space favored to set $\Delta a_{\mu}=0$ to
within 2$\sigma$. These regions are compared with the bounds from
direct detection (dotted) and LEP (dashed). These bounds are shown for
250 (black), 400 (green), and 600 GeV (brown) mediators.}
\end{figure}

\subsection{Dirac DM\label{ssec:dDM}}
\subsubsection{Charged Spin-0 Mediator\label{sssec:mDMts}}
In the case of a charged scalar mediator the interaction Lagrangian 
takes the form
 \begin{align}
\mathcal{L}=\frac{\lambda}{2}\left[\bar{\chi}(1+\gamma_5)\ell
\phi_{\ell}+ \bar{\ell}(1-\gamma_5)\chi\phi^{\dag}_{\ell} \right]
 \end{align} 
 with $\ell$ a lepton and $\phi_{\ell}$ a heavy charged scalar. 

The contribution to $g-2$ can be obtained from \eqref{eq:asf}, and leads to 
the relation
\begin{align}
  \left(\frac{|\lambda|}{m_\phi^2} \right)^2
  &=
  -\frac{96\pi^2 \hat{a}_{\mu}^{S,F}(1-r)^4} 
  {m_{\mu}^2(1-6r^2+3r^4+2r^6-6r^4\ln r^2)}
\end{align} 
 where $r=m_{\chi}/m_\phi$. Although the sign of the contribution implies 
that this theory cannot explain the discrepancy in $a_{\mu}$, it can be 
used to set bounds on the allowed parameter space. These limits are 
compared with those derived from LEP monophoton searches (dashed curves) 
and direct detection (dotted curve) in Fig. \ref{fig:aSF150all}. The 
color of the curves denotes the mass of the mediator, the labelling 
being 250 GeV (black), 400 GeV (green), or 600 GeV (brown).
Additionally, the right-handed slepton bounds from LEP and the LHC
also apply,
and are shown in Fig. \ref{fig:slepsearch}.

The constants 
$c$ and $d$ in \eqref{eq:ddtdampo1} and \eqref{eq:ddtdampo2} were
determined in \cite{Agrawal:2011ze} by matching from the full theory. They lead to
\begin{align}
  \widetilde{\lambda}^{S,F}
  &=
  \frac{\lambda^2\alpha}{16\pi m_{\phi}^2}
  \left(
  1+\frac{2}{3}\sum_{\ell}\ln\frac{m_{\ell}^2}{m_{\phi}^2} 
  .\right)
  \label{eq:ddtdflam}
\end{align} 
 A calculation in the effective theory with only the mediator integrated 
out (performed for this model in \cite{Kopp:2009et}) suffices to determine 
the leading log behavior of $\widetilde{\lambda}$, and is a very good 
approximation to the full result. 

We see from the figure that the direct detection bounds are much 
stronger than those arising from $a_{\mu}$, except in the region of very 
small DM masses, where the limits from monophoton searches are the 
strongest.

\subsubsection{Charged Spin-1 Mediator\label{sssec:dDMtv}}
If the mediator is a charged vector the interaction given by
\begin{align}
  \mathcal{L}
  &=
  \frac{\lambda}{2}
  \left[
  \bar{\chi}\gamma_{\mu}(1+\gamma_{5})\ell V^{\mu}_{\ell} 
  +\bar{\ell}\gamma_{\mu}(1+\gamma_{5})\chi V^{\mu\,\dag}_{\ell} 
  \right]
\end{align}

The contribution to the anomalous magnetic moment of the muon can be 
obtained from \eqref{eq:avf}, and leads to the relation
\begin{align}
  \left(\frac{|\lambda|}{m_V^2} \right)^2
  &=
  \frac{96\pi^2 \hat{a}_{\mu}^{V,F}(1-r)^4} 
  {m_{\mu}^2(10-43r^2+78r^4-49r^6+4r^8+18r^6\ln r^2)}
\end{align}
where $r=m_{\chi}/m_V$. This can be used to determine the region of 
parameter space favored to explain the anomaly in $a_{\mu}$. 

For direct detection, the scattering amplitude in the leading logarithm 
approximation is found to be
 \begin{align}
  \mathcal{M}
  &=
  \frac{\lambda^2\alpha}{12\pi m_V^2}
  \left[
  \ln\frac{m_{\tau}^2}{m_V^2} 
  +\ln\frac{m_{\mu}^2}{m_V^2}
  +\ln\frac{|k|^2}{m_V^2} 
  \right]
  \sum_{q}
  \langle N_f|Q\bar{q}\gamma^{\nu}q|N_{i}\rangle 
  \bar{u}(p_2)\gamma_{\nu}(1+\gamma^{5})u(p_1)
  \label{eq:ddtdamp}
\end{align}
where as before the $\gamma^{\nu}\gamma_5$ terms is velocity
suppressed so we neglect it. We therefore obtain 
\begin{align}
  \widetilde{\lambda}^{V,F}
  &=
  \frac{\lambda^2\alpha}{12\pi m_{V}^2}
  \left[
  \ln\frac{m_{\tau}^2}{m_V^2} 
  +\ln\frac{m_{\mu}^2}{m_V^2}
  +\ln\frac{|k|^2}{m_V^2} 
  \right].
\end{align}

The bounds from 
direct detection (dotted) and LEP monophoton searches (dashed) are 
compared to this favored region in Fig. \ref{fig:aVF150all}. The limits 
are plotted for 250 (black), 400 (green), and 600 GeV (brown) mediators.
We see that values of $\Lambda$ between 165 GeV and 285 GeV 
can explain the anomaly in $a_{\mu}$. However, the LEP and 
direct detection bounds for this range of mediator masses leaves only 
$r$ values smaller than 0.05 unrestricted, so that the DM mass is 
favored to be less than about 10 GeV.

\subsubsection{Neutral Spin-0 Mediator\label{sssec:dDMss}}
\begin{figure}[tp]
\centering
\includegraphics[width=0.45\textwidth]{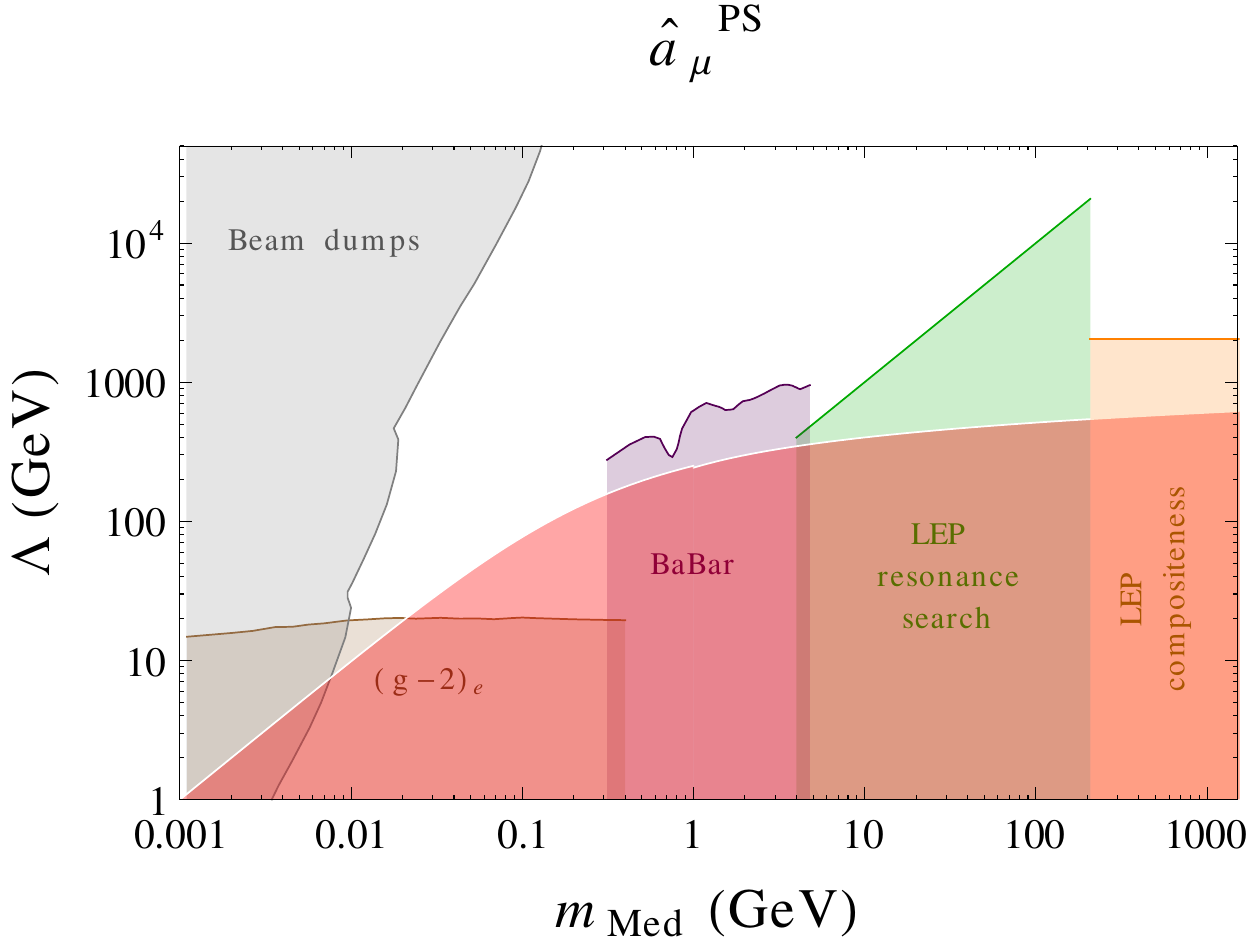}
\caption{\label{fig:aPS1} We show disfavored regions of
parameter space for Dirac fermion DM
with a neutral pseudoscalar mediator as a function of $m_{Med}$.
The shaded exclusion region
 is drawn by requiring the contributions to $a_{\mu}$ 
to be less than 2$\sigma$.
}
\end{figure}

When the scattering of DM off leptons is mediated by a neutral spin-0 
particle $S$ the interaction Lagrangian takes the form
\begin{align}
  \mathcal{L}_S
  &=
  \lambda
  \bar{\mu}\mu S
  &
  \mathcal{L}_{PS}
  &=
  i\lambda\bar{\mu}\gamma^5\mu S
\end{align}
Since the neutral spin-0 mediator does not mix with the photon, the
results obtained for the case of real scalar DM
summarized in Fig~\ref{fig:aS1} continue to apply. The Dirac fermion
DM can also couple to leptons via a pseudoscalar. The corresponding
bounds for this interaction are shown in Fig~\ref{fig:aPS1}.

\begin{figure}[tp]
\centering
\includegraphics[width=0.45\textwidth]{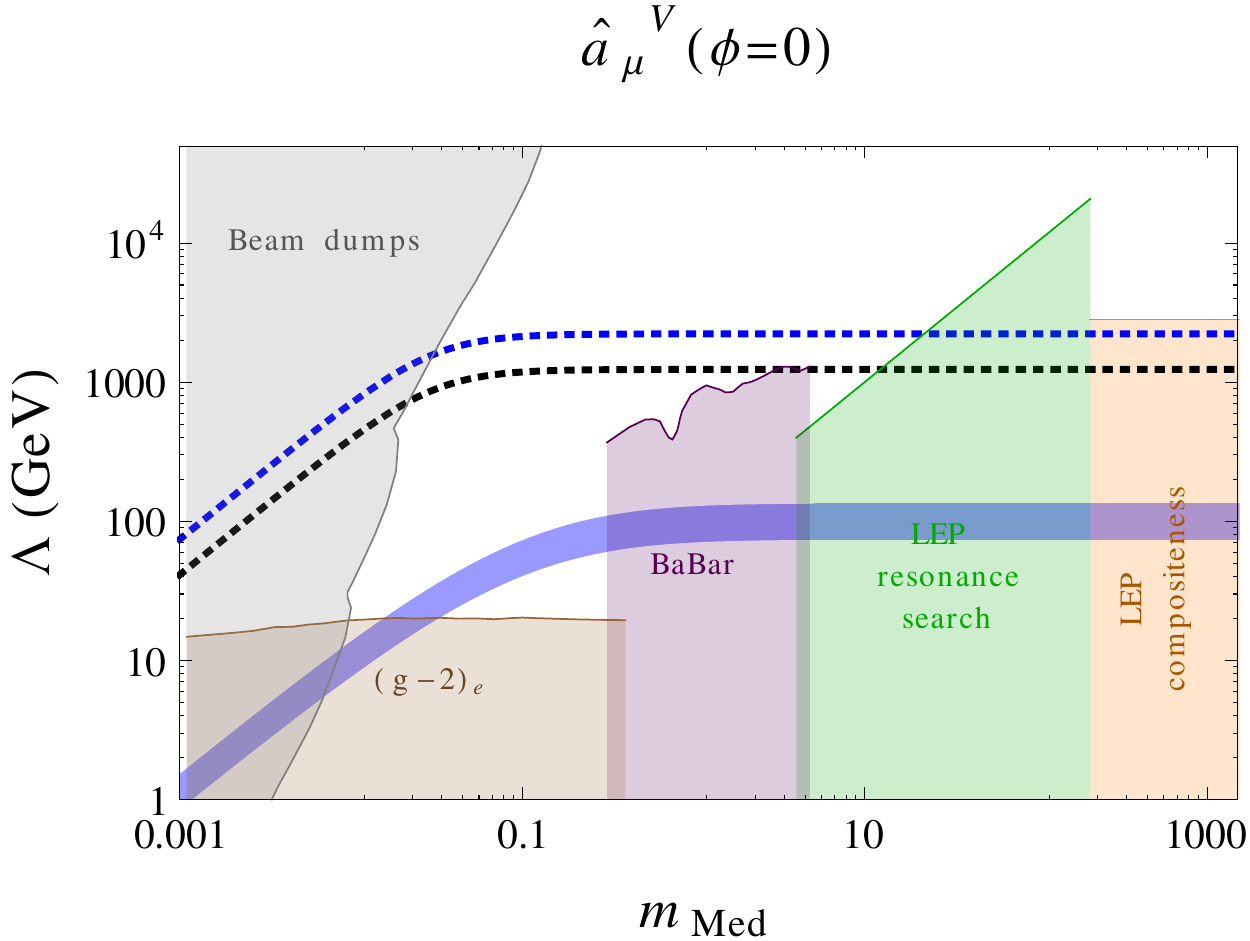}
\caption{We show the region of parameter space
favored by setting $\Delta a_{\mu}=0$ to within 2$\sigma$ (blue
shaded) for Dirac DM
mediated by a neutral vector. The dotted lines show the
direct detection
bounds for a DM mass of 10 GeV (blue) and 90 GeV (black).  }
\label{fig:aV150all}
\end{figure}
\subsubsection{Neutral Spin-1 Mediator}
\label{sssec:dDMsv}
When the interaction is mediated by a neutral spin-1 particle 
the relevant part of the Lagrangian takes the form
\begin{align}
  \mathcal{L}
  &=
  \frac{\lambda_{\chi}}{2}
  \bar{\chi}\gamma^{\nu}
  \left(\cos\theta +\sin\theta\gamma^5 \right)\chi V_{\nu}
  +\frac{\lambda}{2}\bar{\mu}\gamma^{\nu}
  \left(\cos\phi +\sin\phi\gamma^5 \right)\mu V_{\nu}.
\end{align}
 where the mixing angle $\theta$ parametrizes the relative strengths of 
the vector and axial-vector components of the DM interaction. The 
$a_{\mu}$ bounds only depend on $\phi$ and not on $\theta$. The direct detection and monophoton limits, on the 
other hand, depend on both $\theta$ and $\phi$. As before, we pick the 
pure vector ($\theta=\phi=0$) and the pure axial vector 
($\theta=\phi=\pi/2$) cases for illustration. The axial-vector mediator 
does not mix with the photon, and so the direct detection bounds do not 
apply. Consequently, the limits are the same as for the complex scalar DM 
model shown in Fig. \ref{fig:aVdd1}(a).

For the vector-vector interaction, the leading logarithmically
enhanced contribution to the direct detection scattering 
amplitude takes the form
\begin{align}
  \widetilde{\lambda}^V
  &=
  \frac{\lambda^2 \alpha}{3\pi m_V^2}
  \left[\ln\frac{m_{\tau}^2}{\Lambda_V^2} 
  +\ln\frac{m_{\mu}^2}{\Lambda_V^2}
  + \ln\frac{|k|^2}{\Lambda_V^2} \right].
\end{align}
 where $\Lambda_V$ is the scale of new physics below which the mixing 
between the mediator and photon is generated. For concreteness we take 
$\Lambda_V = 1$ TeV. The region favored by the muon $g-2$ 
measurement, and constraints from direct detection and colliders is 
shown in Fig. \ref{fig:aV150all}. For heavy mediator masses ($m_{Med} > 
208$ GeV), LEP compositeness bounds rule out the entire region 
consistent with the muon $g-2$ measurement. For light mediators, direct 
detection constraints are the most severe. These bounds can be avoided 
for DM masses below direct detection experiment sensitivity ($m_{DM} < 
5$ GeV), which leaves a window around for mediator masses between about 
10 MeV and 300 MeV.

\subsection{Real Vector DM\label{ssec:rvecDM}}
\subsubsection{Charged fermion mediator\label{sssec:rvDMt}}
 We conclude our analysis by considering real vector DM candidate 
$\chi^{\nu}$. If the interactions of DM with leptons are mediated by a 
charged fermion we obtain the Lagrangian \begin{align}
  \mathcal{L}
  &=
  \frac{\lambda}{2}
  \bar{\mu}\gamma_{\nu}
  (1+\gamma^5)F\chi^{\nu} +\,\text{h.c.}
\end{align}

The contribution to $a_{\mu}$ for this case can be determined from 
\eqref{eq:afv}. This leads to
 \begin{align}
\left(\frac{|\lambda|}{m_F}\right)^2=-\frac{96\pi^2r^2(1-r)^4 \hat{a}_{\mu}^{F,V}}{m_{\mu}^2(5-14r^2+ 39r^4-38r^6+8r^8+8r^4\ln r^2)}
 \end{align}
with $r=m_{\chi}/m_F$. We see from the sign of the contribution that this
theory cannot explain the observed discrepancy in $a_{\mu}$. 

\begin{figure}[tp]
\centering
\includegraphics[width=0.45\textwidth]{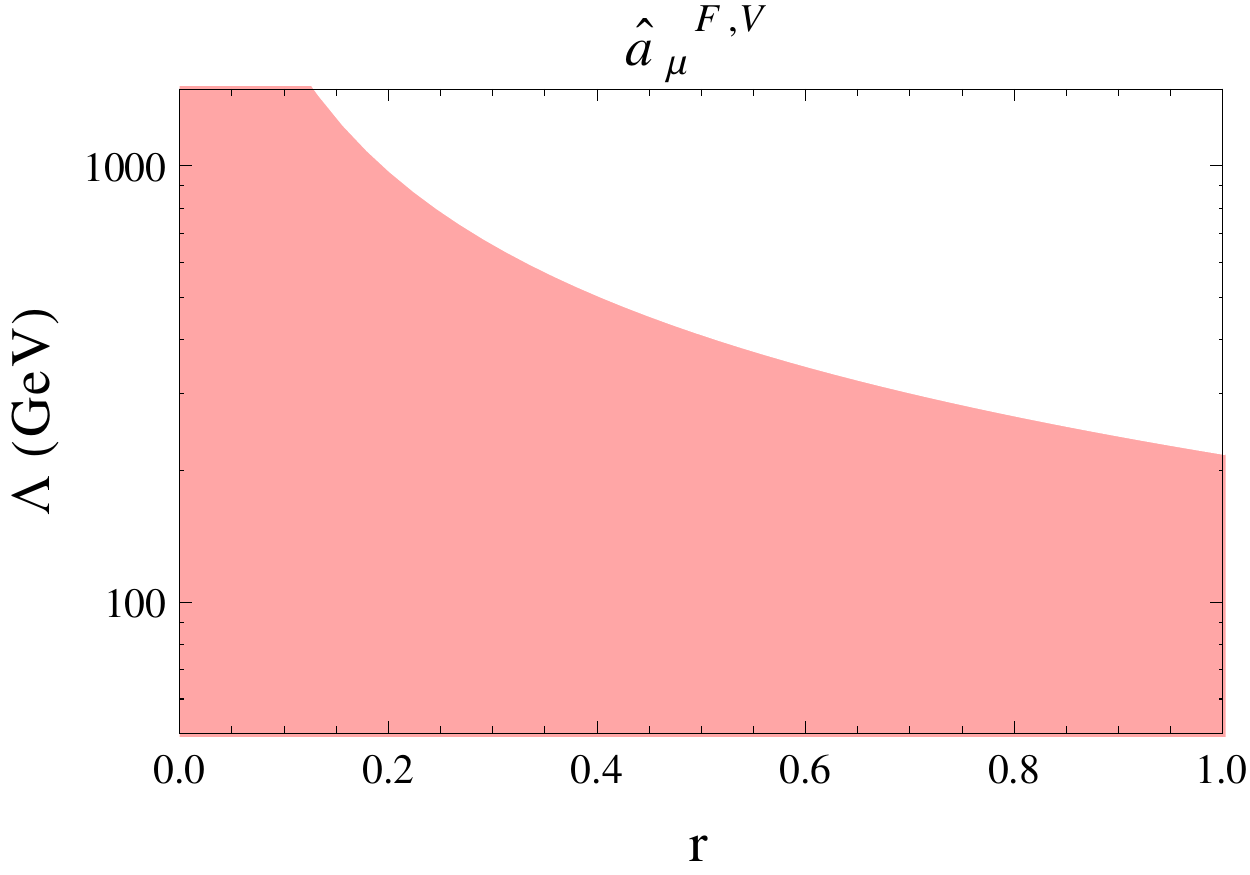}
\caption{\label{fig:RV} We plot the bound set by requiring
the contribution to $a_{\mu}$ by real vector DM mediated by a charged
fermion be smaller than 2$\sigma$, as a function of $r=m_{\text{DM}}/m_{\text{Med}}$. }
\label{fig:aFV2}
\end{figure}

This scenario is not expected to be significantly constrained by the 
current limits from direct detection. The monophoton analyses in the 
literature are also not applicable to this model. We plot the bound 
obtained by from $a_{\mu}$ in Fig. \ref{fig:aFV2} as a function of $r$. We 
find that for small values of $r$ the value of $\Lambda$ is constrained to 
be fairly large. The weakest bound, around 220 GeV, lies close to $r=1$.

\subsubsection{Neutral Spin-0 Mediator\label{sssec:rvDMss}}
When mediation occurs through a neutral spin-0 scalar
$S$ we have the Lagrangian
\begin{align}
  \mathcal{L}_S
  &=
  \lambda\bar{\mu} \mu S
\end{align}
 There have been no monophoton studies performed on this model. There 
are also no significant constraints on this model from direct detection. 
Therefore, the analysis of a spin-0 mediator coupling to real scalar DM 
summarized in Fig.~\ref{fig:aS1} also applies to this case. There are
no $CP$-conserving renormalizable couplings of a pseudoscalar mediator
with real vector DM.

\section{Conclusions \label{sec:conc}}

Within the WIMP paradigm, leptophilic DM offers a simple way to remain
consistent with current collider, direct detection and indirect
detection bounds. These models lead to corrections to the anomalous
magnetic moment of the muon, and can potentially explain the observed
discrepancy between the experimentally measured value and the SM
prediction. From our analysis, it follows that for many of the
simplest leptophilic DM models, a large part of the parameter space
which explains the anomaly is excluded by other experiments.
Nevertheless, there do exist limited regions of parameter space in
these simple theories that can explain the observed anomaly while
remaining consistent with current experimental bounds.

For some theories of leptophilic DM, the contribution to $a_{\mu}$ has the 
wrong sign to explain the anomaly. This can be used to place limits on the 
parameter space of these theories. We find that, at present, these limits 
are competitive with those from collider and direct detection experiments.

\begin{acknowledgments}
  We would like to thank Brian Batell, Patrick Fox, Brian Hamilton 
  and Samuel McDermott for useful discussions.  PA would
  like to acknowledge support by the National Science Foundation under
  Grant No.  PHYS-1066293 and the hospitality of the Aspen Center for
  Physics, where a part of this work was completed. ZC and CV are
  supported by the NSF grant PHY-0968854. Fermilab is operated by
  Fermi Research Alliance, LLC under Contract No. DE-AC02-07CH11359
  with the United States Department of Energy. 
\end{acknowledgments}
{\bf Note added}: While we were completing this paper, we received 
the manuscripts \cite{Bai:2014osa,Freitas:2014pua} 
which overlap with
some of the ideas presented here.

\bibliography{DMg_2bib}

\end{document}